\documentclass[
     ,draft            
  ]
  {aipproc}

\layoutstyle{6x9}

\begin{document}
\title[BPS preons. An overview from the hill of twistor
 appraoch]{BPS preons in supergravity and
 higher spin theories. \\ An overview from the hill of twistor
 appraoch}

\author{Igor A. Bandos}
{address={Departamento de F\'{\i}sica Te\'orica, Univ.~de Valencia and IFIC
(CSIC-UVEG), 46100-Burjassot (Valencia), Spain \\ Institute for Theoretical Physics,
NSC ``Kharkov Institute of Physics  and Technology'',  UA61108, Kharkiv, Ukraine} }

\begin{abstract}
We review briefly the notion of BPS preons, first introduced in 11--dimensional context
as hypothetical constituents of M--theory, in its generalization to arbitrary
dimensions and emphasizing   the relation with twistor approach. In particular, the use
of a ''twistor--like'' definition of BPS preon (almost) allows us to remove
supersymmetry arguments from the discussion of the relation of the preons with higher
spin theories and also of the treatment of  BPS preons as constituents. We turn to the
supersymmetry in the second part of this contribution, where we complete the algebraic
discussion with supersymmetric arguments based on the M--algebra (generalized
Poincar\'e superalgebra), discuss the possible generalization of BPS preons related to
the $osp(1|n)$ (generalized AdS)  superalgebra, review a twistor--like
$\kappa$--symmetric superparticle in tensorial superspace, which provides a point--like
dynamical model for BPS preon, and the r\^ole of BPS preons in the analysis of
supergravity solutions. Finally  we describe resent results on the concise superfield
description of the higher spin field equations and on superfield supergravity in
tensorial superspaces.

\medskip
{\bf 14/01/2005. V2: References added, citations completed, 28/01/2005}

\end{abstract}

\maketitle


\section{Introduction }

Twistor theory \cite{Pen,PenTP} and twistor--like methods, which are the main subjects
of this Max Born symposium,  are becoming now increasingly popular in the light of the
work of \cite{W03,W04} on the twistor string description of the Yang--Mills scattering
amplitudes \cite{Nair}. This  can be considered as a significant progress towards a
realization of the Penrose ``twistor programme'' \cite{PenTP} aimed to describe nature
in terms of twistor space rather than spacetime.

The subject of this contribution is the notion of BPS preons, introduced in
\cite{BPS01} in an M--theoretical context, but allowing for an easy 'generalization' to
other dimensions (see \cite{B02,30/32} and also \cite{Rubek}). In M-theory the BPS
preons appeared as its (hypothetical) constituents \cite{BPS01}; a search 31/32
supersymmetric solutions of $D=11$ supergravity, which would describe BPS preons, can
be witnessed \cite{Duff03,Hull03,BPS03}. In some other dimensions, namely in  $D= 4, 6$
and $10$, the notion of BPS preons  are related with higher spin theory (see
\cite{BLS99,V01s,V01c,30/32,BPS03,Dima03,Misha0304b,BPST04,BBdAST04}).

As it was noticed already in \cite{BPS01} the notion of BPS preons  is related to the
twistor approach \cite{Pen} and its very simple orthosymplectic ``generalization''
\cite{BL98} (hence the ``twistorial constituents'' name in the  title of \cite{BPS01}).
The discussion of this relation allows us to define the BPS preon in a simple and
suggestive way, with a minimal use of supersymmetry. This observation suggests the
following structure of this contribution.

 We begin in Sec. I by a brief review of the known properties of twistor approach,
 massless particle mechanics
and their supersymmetric generalization in the form which is useful to define and to
discuss the properties of BPS preons.  In Sec. II  we present a purely bosonic
definition of the of BPS preon  \cite{BPS01} and discuss  their properties (almost)
without using (more precisely, with a minimal reference to) supersymmetry. In this
framework we review, in particular, the r\^ole of BPS preons as constituents (of
M--theory for $D=11$)  and the relation of BPS preon with higher spin theories. To
establish this relation we use the point--like model for BPS preon provided by the
twistor--like (super)particle model in tensorial (super)space; interestingly enough,
this model had been proposed in \cite{BL98} before the notion of BPS preons was
introduced in \cite{BPS01}. The same can be said (at least up to some extent, see
\cite{BLS99} and \cite{V01s,V01c,Dima03,BPST04,BBdAST04}) on the relation of this model
with $D=4,6,10$ higher spin theories. We use here the BPS preon notion to discuss these
issues as it provides a universal framework allowing to discuss the higher spin theory
in $D=4,6,10$ and (some issues of) M--theory in the same term.  The discussion of Sec.
II is completed by supersymmetry arguments in  Sec. III where we start form M--algebra,
discuss the r\^ole of BPS preons in the classification of the BPS state \cite{BPS01}
and in the analysis of the supergravity solitons \cite{BPS03}, review the
$\kappa$--symmetry of the ``preonic superparticle'' model \cite{BL98} and its relation
with preserved supersymmetry. We finish in Sec. IV by describing  recent results on the
superfield description of the tower of all possible conformal massless higher spin
equations in $D=4,6,10$ and on supergravity in tensorial superspaces, which might be
relevant both in the search for a selfconsistent supersymmetric higher spin interaction
and for M--theoretical applications.

\section{I. Preliminaries. Twistor approach, massless particle and superparticle in $D=4$. }

This section contains a review of known issues on $D=4$ twistors and supertwistors  and
their relation to massless particle and superparticle mechanics \cite{Ferber} (see also
e.g. \cite{Dima89,Dima} and a more recent \cite{FZ,JdA+L+C}) in a  form convenient for
the discussion on BPS preons.

\subsection{I.1. Cartan--Penrose representation and Penrose
correspondence}

Let us begin by writing two basic relation of the original Penrose twistor approach in
$D=4$ \cite{Pen}. One is the Cartan--Penrose representation  for  a real light--like
vector, {\it e.g.} the momentum of massless particle,
\begin{eqnarray}
\label{P=ll4D} p_{A\dot{A}} &:=& p_{a} \sigma^a_{A\dot{A}} =
\lambda_A \bar{\lambda}_{\dot{A}} \; , \qquad \Leftrightarrow \qquad
p_ap^a=0 \;
\end{eqnarray}
($a=0, 1,2,3$, $A=1,2$ and  $\dot{A}=1,2$ are Weyl spinor indices
and $ \sigma^a_{A\dot{A}}$ are relativistic Pauli matrices).
Another is the famous Penrose correspondence,
\begin{eqnarray}
\label{m=xl} \mu^{\dot{A}} =  x^{\dot{A}A} \lambda_A  &:=& {1\over
2} x^{a} \sigma_a^{\dot{A}A}\lambda_A \; .
\end{eqnarray}
For a fixed   $x^a$ (real or complex), Eq. (\ref{m=xl}) is a
homogeneous linear equation for the coordinates $Y_{0\alpha}=
(\mu^{\dot{A}}\,, \lambda_A)$ of the complex space
$\mathrm{C}^{4}$. Imposing the topological restriction
$Y_{0\alpha} \not= (0,0,0,0)$ (that is passing from
$\mathrm{C}^{4}$ to $\mathrm{C}^{4}- \{ 0\}$) and using the
scaling symmetry $(\mu^{\dot{A}} , \lambda_A)\mapsto (z
\mu^{\dot{A}} , z\lambda_A)$ as an
 identification relation, one can treat $Y_{0\alpha}= (\mu^{\dot{A}} , \lambda_A)$
as  {\it homogeneous}  coordinates of {\it the projective twistor
space} $\mathrm{CP}^{3}$
 \cite{Pen,PenTP}.
Thus, as usually stated in twistor approach, Eq.  (\ref{m=xl}) describes a
correspondence the space of  light--like lines in spacetime (which can be identified
with the celestian sphere $S^{2}$) and the set of all surfaces in the projective
twistor space $\mathbf{CP}^{3}$ (``curves of genus zero and degree one'' \cite{W03})
which is isomorphic to $\mathbf{CP}^1$ (in this sense one can say that the Penrose
correspondence illustrates  the known identity $S^{2}= \mathbf{CP}^1$).

To understand that the correspondence involves light--like lines rather then points
$x^a$ of the Minkowski spacetime $\mathrm{M}^4$, one  notices the  symmetry of Eq.
(\ref{m=xl}) under
\begin{eqnarray} \label{b-sym}
 \delta x^{\dot{A}A} = b\, \lambda^{\dot{A}}\lambda^A \; ,  \qquad
\end{eqnarray}
which is usually called $b$--symmetry. The presence of an arbitrary parameter $b$ as a
coefficient for light--like vector $\lambda^{\dot{A}}\lambda^A$ implies that the orbit
of the b--symmetry transformations (\ref{b-sym}) is the light--like line
$\hat{x}^{\dot{A}A}(b) = x^{\dot{A}A} + b \lambda^A \bar{\lambda} ^{\dot{A}}$.

Let us notice that Eq. (\ref{m=xl}) with real $x^a$ is the general solution of the
single real equation for the twistor variables. This is usually called {\it helicity
constraint}, and  reads
\begin{eqnarray}
\label{lyl=0} {\mathcal{S}}= \bar{\mu}^{\dot{A}}
\bar{\lambda}_{\dot{A}} - {\mu}^{{A}} {\lambda}_{{A}} = 0  \; .
\end{eqnarray}
If one substitute the complex  vector $x_L^a$ (the non--Hermitian
 $x_L^{\dot{A}A}$) for the real $x^a$
in (\ref{m=xl}), thus studying the Penrose correspondence in the complexified Minkowski
spacetime $\mathrm{CM}^4$ (see {\it e.g.} \cite{Pen}),
\begin{equation}\label{m=xLl0}
\mu^{\dot{A}}=x_L^{\dot{A}A} \lambda_A\; , \qquad x_L^{\dot{A}A}=
x^{\dot{A}A} + iy^{\dot{A}A} \quad ,
\end{equation}
one finds, instead of (\ref{lyl=0}),  ${\mathcal{S}}= \bar{\mu}^{\dot{A}}
\bar{\lambda}_{\dot{A}} - {\mu}^{{A}} {\lambda}_{{A}} = 2i
\bar{\lambda}_{\dot{A}}y^{\dot{A}A}\lambda_A$, where $y^{\dot{A}A}:=   1/2i(
x_L^{\dot{A}A}- (x_L^{\dot{A}A})^*)$ is the imaginary part of $x_L^{\dot{A}A}$. The
correspondence with the complexified Minkowski spacetime $\mathrm{CM}^4$ can be used,
in particular, to describe fields of nonzero helicity \cite{Ferber}: to this end one
sets ${\mathcal{S}}= \bar{\mu}^{\dot{A}} \bar{\lambda}_{\dot{A}} - {\mu}^{{A}}
{\lambda}_{{A}} = 2is$ with some half--integer $s$. \footnote{ Notice that nonzero
helicities can appear as a result of the ordering ambiguity after quantization of the
massless particle; the quantum consideration also indicates the quantization of
helicity $s$ in the units of $\hbar /2$, see \cite{Balachandran,EiS,B90} and refs.
therein.} Notice also that the one--parametric $b$--symmetry (\ref{b-sym}) in the case
of complex $x_L^a$ is replaced by the complex--spinor--parametric symmetry
$x^{\dot{A}A} + u^{\dot{A}}\lambda^A$ of (\ref{m=xLl0}). This allows to gauge away all
the imaginary part $y^a$ of
 $x_L^a$ except for the one enclosed in the contraction
$\lambda_Ay^{\dot{A}A}\bar{\lambda}_{\dot{A}}\equiv 1/2i {\mathcal{S}}_c$. Thus the
helicity constraint with nonvanishing {\it r.h.s.}, ${\mathcal{S}}_c= 2is$ may be used,
together with gauge fixing, to define $y^a = \Im m\, ( x_L^a)$ completely:
$y^{\dot{A}A}= s \, w^A \bar{w}^{\dot{A}}$ where $w^A\lambda_A = 1$ and
$\bar{w}^{\dot{A}}=(w^A)^\ast$.

An important observation is that, in distinction to (\ref{m=xl}), the equation
(\ref{m=xLl0}) with an {\it arbitrary complex} $x_L^a$ does not restrict the twistor
$Y_{0\, \alpha}= (\mu^{\dot{A}}, \lambda_A)$ at all but rather provides a possibility
to change a set of basic variables form the set of two spinor $(\mu^{\dot{A}},
\lambda_A)$ to one spinor $\lambda_A$ and a complex vector $x_L^{\dot{A}A}=x^{\dot{A}A}
+ iy^{\dot{A}A}$ defined modulo the (gauge) transformations $x_L^{\dot{A}A}\mapsto
x_L^{\dot{A}A}+ u^{\dot{A}} \lambda^A$. This can be treated  as a reason for the
existence of the formulation of bosonic higher spin theory with an auxiliary vector
variable whose $AdS$ version allows for a  nontrivial interaction \cite{Misha0304b}.

\subsection{I.2.Twistors and massless particle in $D=4$ Minkowski
spacetime}

The one parametric $b$ symmetry (\ref{b-sym}), which is the invariance of Eq.
(\ref{m=xl}), can be identified with  the gauge  symmetry of the massless particle
action in its Ferber--Schirafuji form \cite{Ferber}
\begin{eqnarray}\label{Sb4D=ll}
S_0= \int_{W^1} d\hat{x}^{\dot{A}A} \hat{\lambda}_A
\hat{\bar{\lambda}}_{\dot{A}} \equiv \; \int d\tau \;
\partial_\tau\,  \hat{x}^a(\tau) \; \tilde{\sigma}_a^{\dot{A}A}\;
\hat{\lambda}_A\hat{\bar{\lambda}}_{\dot{A}}(\tau) \; .
\end{eqnarray}
A simple way to obtain this ''twistor--like'' action is to start
with the first--order form of the Brink--Schwarz formulation of
the massless particle action, $S_{0\, BS}= {1\over 2} \int_{W^1}
(\hat{p}_{A\dot{A}}\, d\hat{x}^{\dot{A}A} +  {1\over 2} d\tau \, e
 \, \hat{p}_{A\dot{A}}\hat{p}^{\dot{A}A} )$,  and to
 substitute the general solution
\begin{eqnarray}\label{p0=ll4D}
\hat{p}_{A\dot{A}} = \hat{\lambda}_{A}
\hat{\bar{\lambda}}_{{\dot{A}}} \qquad (\Leftrightarrow \qquad
\hat{p}_a\hat{p}^a\equiv {1/2}
\hat{p}_{A\dot{A}}\hat{p}^{\dot{A}A} =0 \; ) \;
\end{eqnarray}
of the algebraic equation of motion  $\; {\delta S_{BS}\over \delta e}= {1\over 2} p_a
p^a =0\,$ for an arbitrary $\hat{p}_{A\dot{A}}$ in $S_{0\; BS}$. One can also obtain
the $b$--symmetry (see (\ref{b-sym}) of the action (\ref{Sb4D=ll}),
\begin{eqnarray}\label{bsym-xll}
 \delta_b \hat{x}^{\dot{A}A} = b \hat{\lambda}^A
\hat{\bar{\lambda}}{}^{\dot{A}}  \; , \qquad \delta_b
\hat{\lambda}^A = 0
\end{eqnarray}
by substituting the Cartan--Penrose representation (\ref{P=ll4D})
for the light--like $\hat{p}_a$, given by Eq.  (\ref{p0=ll4D}),
 in the gauge symmetry
$\delta \hat{x}^{a}= b\, \hat{p}^{a}$ of the action
$S_{BS}$\footnote{This symmetry also includes  $\delta e=
\partial_\tau b(\tau)$, $\delta \hat{p}_a=0$ and
constitutes a ``variational version'' of  worldline reparametrization.}.

The massless particle action (\ref{Sb4D=ll}) provides a simple way to see the relation
among the two basic formulae of the twistor approach, namely among the Cartan--Penrose
representation of Eq. (\ref{P=ll4D}) and the Penrose correspondence relation
(\ref{m=xl}). First one notices that the Hamiltonian formalism for the action
(\ref{Sb4D=ll}) \cite{Ferber} reproduces  the worldline version (\ref{p0=ll4D}) of Eq.
(\ref{P=ll4D}) as a primary constraint (see \cite{Dirac}). Secondly, the above
observation that the action possesses  the same b--symmetry as the Penrose
correspondence (\ref{m=xl}) (see Eqs. (\ref{bsym-xll}) and (\ref{b-sym})) suggests that
(\ref{Sb4D=ll})   should also reproduce the worldline version of Eq. (\ref{m=xl}),
\begin{equation}\label{hm=hxl0}
\hat{\mu}^{\dot{A}}=\hat{x}^{\dot{A}A} \hat{\lambda}_A\; .
\qquad
\end{equation}
This is indeed the case. Using the Leibnitz rule ($d\hat{x} \; \hat{\lambda}
\hat{\bar{\lambda}} \equiv d(\hat{x} \hat{\lambda})\, \hat{\bar{\lambda}} - \hat{x}
\hat{\bar{\lambda}}\, d\hat{\lambda}$) one can write the action (\ref{Sb4D=ll}) in the
form (see \cite{Ferber})
\begin{eqnarray}\label{Sb4D=mdl}
& S_0= \int_{W^1} (d\bar{\mu}^{\dot{A}}
\hat{\bar{\lambda}}_{\dot{A}} - \hat{\mu}^{{A}} d\hat{\lambda}_A)
\; , \qquad  \hat{\bar{\mu}}^{\dot{A}}
\hat{\bar{\lambda}}_{\dot{A}} -\hat{\mu}^{{A}} \hat{\lambda}_A
 =0 \; ,
\end{eqnarray}
where $\hat{\bar{\mu}}{}^{\dot{A}}$ is defined by Eq. (\ref{hm=hxl0}); as
 (\ref{hm=hxl0}) is the general solution of the helicity
constraints (\ref{lyl=0}), one can, alternatively, consider the
 twistor variables $\hat{Y}_{0\alpha}=(\hat{\bar{\mu}}{}^{\dot{A}}, \hat{\lambda}_A)$ to be
subject to the helicity constraint (\ref{lyl=0}) [as it is written
in (\ref{Sb4D=mdl})] and omit any reference on the spacetime
coordinates.
 Taking the second point of view one finds that just the constraint (\ref{lyl=0})
reduces the imaginary part of the action (\ref{Sb4D=mdl}) to a total  derivative. This
constraint  can be also incorporated into the action with a Lagrange multiplier $\Xi
(\tau)$,
\begin{eqnarray}\label{Sb4D=mdl+c}
S= \int_{W^1} (d\hat{\bar{\mu}}^{\dot{A}} \,
\hat{\bar{\lambda}}_{\dot{A}} - \hat{\mu}^{{A}} d\hat{\lambda}_A) +
\int_{W^1}\,  d\tau\;  \Xi (\tau)\,  (\hat{\bar{\mu}}^{\dot{A}}
\hat{\bar{\lambda}}_{\dot{A}} -\hat{\mu}^{{A}} \hat{\lambda}_A) \; .
\end{eqnarray}

\subsection{I.3. Supersymmetry: massless superparticle and
supertwistors}

The supersymmetric generalization of the action (\ref{Sb4D=ll}) can be obtained {\it
e.g.} starting with the first order form of the  Brink--Schwarz superparticle action
and using there the general solution (\ref{p0=ll4D}) of the mass shell constraints
$p^2=0$ (see Sec.I.2). It reads \cite{Ferber}
\begin{eqnarray}\label{S4D=ll}
S= \int_{W^1} \hat{\Pi}^{\dot{A}A} \hat{\lambda}_A
\hat{\bar{\lambda}}_{\dot{A}} \equiv \; \int d\tau \;
 \, \hat{\lambda}_A \tilde{\sigma}_a^{\dot{A}A}
\hat{\bar{\lambda}}_{\dot{A}}\;  \hat{\Pi}_\tau^a(\tau) , \;
\end{eqnarray}
where $\hat{\Pi}^{\dot{A}A}=d\tau \Pi_\tau^{\dot{A}A}$ is the pull--back to the
particle worldline $W^1$ of the Volkov--Akulov one--form
\begin{eqnarray}\label{Pi4D}
\Pi^a = d x^a - i d\theta_i \sigma^a \bar{\theta}^i + i \theta_i
\sigma^a d\bar{\theta}^i \quad \Leftrightarrow \quad
\Pi^{\dot{A}A}= dx^{\dot{A}A} - id\theta^A_i \,
\bar{\theta}^{\dot{A}i} + i\theta^A_i \, d\bar{\theta}^{\dot{A}i}
\,  \quad
\end{eqnarray}
for the $D=4$ N--extended superspace $\Sigma^{(4|4N)}$ with the
local coordinates
\begin{eqnarray}\label{4DSSP}
\Sigma^{(4|4N)}&: \quad z^M= (x^a\, , \, \theta_i^{A}\, , \,
\bar{\theta}^{\dot{A}i}) \, ; \quad a=0, 1, 2, 3\; , \;\; A=1,2 \;
, \; \dot{A}=1,2\; , \;\; i=1, \ldots , N\, . \quad
\end{eqnarray}
 The (N--extended) global  supersymmetry
transformations which leave (\ref{Pi4D}) invariant are
\begin{eqnarray}\label{4Dsusy}
\delta x^a = - i \theta_i \sigma^a \bar{\varepsilon}^i + i
\varepsilon_i \sigma^a \bar{\theta}^i \; , \qquad \delta
\theta^{{A}}_i = \varepsilon_i \; , \qquad
 \delta \bar{\theta}^{\dot{A}i} = \bar{\varepsilon}^{\dot{A}i} \; .
\end{eqnarray}
As in the purely bosonic case, using the Leibnitz rule one can write the action
(\ref{S4D=ll}) in the form ({\it cf.} (\ref{Sb4D=mdl}))
\begin{eqnarray}\label{Ss4D=mdl}
S= \int_{W^1} (d\hat{\bar{\mu}}^{\dot{A}}
\hat{\bar{\lambda}}_{\dot{A}} -  \hat{\mu}^{{A}} d\hat{\lambda}_A -
2id\hat{\eta}_i \, \hat{\bar{\eta}}^i)=
 \int_{W^1} \, d\hat{\Upsilon}_\Lambda
\hat{\bar{\Upsilon}}^{\Lambda} \equiv  \, \int_{W^1} \,
d\Upsilon_\Lambda \Omega^{\Lambda \Pi}\, (\Upsilon_\Pi)^\ast \, .
\end{eqnarray}
Here (the pull--backs of) the components of supertwistor
\begin{eqnarray}\label{Upsilon=}
\Upsilon_\Lambda := (Y_{0\alpha} , \eta_i) = (\mu^{\dot{A}}\, , \,
\lambda_A \, , \,\; \eta_i) \quad \left( \bar{\Upsilon}^{\Lambda}
:=
 \,  \Omega^{\Lambda \Pi}\,
(\Upsilon_\Lambda)^\ast \, \equiv ( \bar{\lambda}_{\dot{A}}\, , \,
\quad \bar{\mu}^A \, , \,  -2i \bar{\eta}^i )^T \right)
\end{eqnarray}
 are related to the coordinates (\ref{4DSSP})
(coordinate functions in (\ref{S4D=ll})) by the following
supersymmetric generalization of the Penrose correspondence
relation (\ref{m=xl}) \cite{Ferber}
\begin{eqnarray}
\label{m=xLl} && \cases{\mu^{\dot{A}} =  x_L^{\dot{A}A} \lambda_A
\; := {1\over 2} x_L^{a} \sigma_a^{\dot{A}A}\lambda_A \; :=
(x^{\dot{A}A} + i \theta_i^A \bar{\theta}^{\dot{A}i}) \, \lambda_A
\; , \cr  \qquad  \eta_i = \theta_i^A  \lambda_A \; . }
\end{eqnarray}
These expressions for the supertwistor gives the general solution of the superhelicity
constraint
\begin{eqnarray}\label{S-U(1)}
{\mathcal S} := \hat{\bar{\mu}}^{\dot{A}}
\hat{\bar{\lambda}}_{\dot{A}} -\hat{\mu}^{{A}} \hat{\lambda}_A - i
\eta_i \, \bar{\eta}^i \equiv \Upsilon_\Lambda
\bar{\Upsilon}^{\Lambda} \equiv \Upsilon_\Lambda \Omega^{\Lambda
\Pi}\, (\Upsilon_\Pi)^\ast
 = 0  \; ,
\end{eqnarray}
in which (as well as in (\ref{Ss4D=mdl})) $\Omega^{\Lambda \Pi}$
\begin{eqnarray}\label{Omega}
& \Omega^{\Lambda \Pi} := \left(\matrix{ \Omega^{\alpha\beta} & 0
\cr 0 & - 2i \delta_i{}^j }\right)\; =
  \, \left(\matrix{ 0 &
\delta_{\dot{A}}{}^{\dot{B}} & 0 \cr - \delta^A{}_B & 0 & 0 \cr 0
& 0 & - 2i \delta_i{}^j  }\right)\qquad
\end{eqnarray}
is the $SU(2,2|N)$ invariant matrix. Such an observation allows one to write the
superparticle action in an equivalent form ({\it cf.} (\ref{Sb4D=mdl+c}))
\begin{eqnarray}\label{Ss4D=mdl+c}
S&= \int_{W^1} (d\hat{\bar{\mu}}^{\dot{A}}
\hat{\bar{\lambda}}_{\dot{A}} -  \hat{\mu}^{{A}} d\hat{\lambda}_A
- 2id\eta_i \, \bar{\eta}^i) + \int_{W^1}\,  d\tau\;  \Xi (\tau)\,
(\hat{\bar{\mu}}^{\dot{A}} \hat{\bar{\lambda}}_{\dot{A}}
-\hat{\mu}^{{A}} \hat{\lambda}_A - 2i \eta_i \, \bar{\eta}^i )
\equiv  \nonumber \\ &   \equiv \int_{W^1}\, d\Upsilon_\Lambda
\Omega^{\Lambda \Pi}\, (\Upsilon_\Pi)^\ast + \int_{W^1}\,  d\tau\;
\Xi (\tau)\, \Upsilon_\Lambda \Omega^{\Lambda \Pi}\,
(\Upsilon_\Pi)^\ast  \; .
\end{eqnarray}
In this form the $SU(2,2|N)$ symmetry of the superparticle action becomes manifest. The
action  (\ref{Ss4D=mdl+c}), incorporating also the constraint (\ref{S-U(1)}) with the
Lagrange multiplier $\Xi$, involves only one constant tensor $\Omega^{\Lambda \Pi}= -
(-)^{\Lambda + \Pi}\Omega^{\Pi\Lambda}$, Eq. (\ref{Omega}), and the invariance of such
a tensor is the defining property of the $SU(2,2|N)$ supergroup.

In relation with the evident equivalence of the action (\ref{Ss4D=mdl+c}) (or
(\ref{Ss4D=mdl}))  with (\ref{S4D=ll}) one can ask  questions about degrees of freedom.
In particular, the action (\ref{S4D=ll})  contains $4N$ fermionic fields (coordinate
functions) $\theta^\alpha$ while (\ref{Ss4D=mdl+c}) (or (\ref{Ss4D=mdl})) involves $2N$
fermionic $\eta_i, \bar{\eta}^i$. This seeming mismatch indicates the presence of $2N$
local fermionic gauge symmetries in the action (\ref{S4D=ll}). These has the form
\begin{eqnarray}\label{k-sym4D}
\delta_\kappa \hat{x}^{\dot{A}A} (\tau) &=& i \, \kappa \,
\lambda^A
 \bar{\theta}^{\dot{A}} + i \, \bar{\kappa} \, \theta^A \,
\bar{\lambda}^{\dot{A}}= i \, \delta_\kappa \, \theta^A
 \bar{\theta}^{\dot{A}} (\tau) - i  \, \theta^A \, \delta_\kappa
\bar{\theta}^{\dot{A}} (\tau) \qquad \\  \delta_\kappa \, \theta^A
&=& \kappa (\tau) \, \lambda^A \; , \qquad \delta_\kappa \,
\bar{\theta}^{\dot{A}}(\tau) = \bar{\kappa} (\tau) \,
\bar{\lambda}^{\dot{A}}(\tau) \; \qquad
\end{eqnarray}
and provide an irreducible form (see \cite{stv,BZ,BZstr,Dima} and refs therein) of the
seminal $\kappa$--symmetry \cite{AL82} of the Brink--Schwarz superparticle which can be
defined by $\delta_\kappa \bar{\theta}^{\dot{A}} \Pi_{\tau\; A\dot{A}}=0$ and
$\;i_\kappa \Pi^{\dot{A}A} := \delta_\kappa \hat{x}^{\dot{A}A} - i \, \delta_\kappa \,
\theta^A
 \bar{\theta}^{\dot{A}}  + i  \, \theta^A \, \delta_\kappa
\bar{\theta}^{\dot{A}}=0$.

\section{II. BPS preons without supersymmetry }

\subsection{II.1. BPS preons and generalized Cartan--Penrose representation}

In this section we present the definition of the BPS preon from \cite{BPS01} (see also
\cite{30/32,BPS03}) in its bosonic form which makes transparent the relation with the
twistor approach.

$\diamondsuit$  The {\it BPS preon  state}  can be characterized
by one
 bosonic spinor $\lambda_\alpha$,
\begin{eqnarray}\label{pr=lb}
{ \vert BPS\; preon \rangle = \vert \, \lambda_\alpha  \rangle \; } , \qquad \alpha=1,
\ldots , n \; ,
\end{eqnarray}
and is an eigenvector of the generalized
 momentum operator $P_{\alpha\beta}=P_{\beta\alpha}$ for the
 eigenvalue $\lambda_\alpha \, \lambda_\beta$ determined by the above mentioned spinor
 $\lambda_\alpha$,
 \begin{eqnarray}\label{(P-ll)pr}
 P_{\alpha\beta} \vert \, \lambda_\alpha  \rangle = \lambda_\alpha \, \lambda_\beta \vert \,
 \lambda_\alpha  \rangle \; .
 \end{eqnarray}
  $\diamondsuit$ In the original M--theoretic context of \cite{BPS01} $\alpha= 1,
 \ldots , 32$ is the Majorana--Weyl spinor index of $SO(1,10)$  ($D=11=1+10$), but
 a generalization for $\alpha=1, \ldots , n$ with other $n=2^k$ allowing treatment
 as Majorana or pseudo--Majorana spinors of $SO(t,D-t)$ with other $D$
 is straightforward.\footnote{The cases of $n\not= 2^k$ $\alpha$ allows for a
 treatment as multispinor index (a set of spinor indices); {\it e.g.} for the  odd values of
 $n$, $\lambda_\alpha$ one can treat $\alpha= 1,
 \ldots , n$  as a set of $n$ one--valued Majorana--Weyl spinors in $D=2$. }

 In supersymmetric theory, where the generalized momentum
is defined by the anticommutator of fermionic charges, $P_{\alpha\beta}= \{ Q_\alpha \,
Q_\beta \}$, the above definition implies (see \cite{BPS01,30/32}) that the BPS preon
state {\it $\vert BPS\; preon \rangle= \vert \lambda_\alpha \rangle$ preserves all but
one supersymmetries generated by $Q_\alpha$} with $\alpha= 1, \ldots , n$. Hence
another notation for the preonic state is $\vert BPS\; preon \rangle= \vert  BPS\;
(n-1) \rangle$ reflecting the number of preserved supersymmetries; in the M--theoretic
$n=32$ case this is $\vert BPS\; preon \rangle= \vert BPS\; 31 \rangle$. This notation,
however, can be understood also without references on supersymmetry, as we will see in
a moment.

\subsection{II.2. BPS preons as fundamental constituents}

The above definition of the BPS preon is based on the eigenvalue problem for the
generalized momentum operator $P_{\alpha\beta}$ and, hence, assumes  that different
components of $P_{\alpha\beta}$ can be diagonalized simultaneously. This is the case
when they are commuting,
\begin{eqnarray}\label{PP=0b}
{} [ P_{\alpha\beta}\, , \, P_{\gamma\delta} \, ] = 0 \; .
 \end{eqnarray}
The general eigenvector $\vert p_{\alpha\beta} \rangle $ of the
Abelian $P_{\alpha\beta}$,
\begin{eqnarray}\label{PBPSp} P_{\alpha \beta}
\vert p_{\alpha\beta} \rangle = p_{\alpha\beta} \vert
p_{\alpha\beta} \rangle \; ,
\end{eqnarray}
is characterized by an eigenvalue matrix $p_{\alpha\beta}$.  One can rise the question
how to classify these states. Such a classification problem looks much less academic in
a supersymmetric context where (see Sec. III) it is equivalent to the search for a
classification of the BPS states (M--theory BPS states for $n=32$) \cite{BPS01}.

The Abelian algebra of the generalized momenta, Eq. (\ref{PP=0b}), possesses a manifest
$GL(n)$ symmetry. The only property  of $\vert p_{\alpha\beta} \rangle $ states which
is invariant under this $GL(n)$ symmetry is the rank, $\mathrm{rank}
(p_{\alpha\beta})$, of the eigenvalue matrix
 $p_{\alpha\beta}$. Let us denote the matrix of rank $(n-k)$ by $p^{\vert k
\rangle}_{\alpha\beta}$, $\mathrm{rank} (p^{\vert k
\rangle}_{\alpha\beta}):=  (n-k)$ ($(32-k)$ in the M--theoretical
case) and the state with the eigenvalue matrix $p^{\vert k
\rangle}_{\alpha\beta}$ by $ \vert BPS\; , \, p^{\vert k
\rangle}_{\alpha\beta}\, \rangle$ or, shortly, $ \vert BPS\; k \,
\rangle = \vert \; k \; \rangle$,
\begin{eqnarray}\label{kBPSp0}
P_{\alpha \beta} \vert k \, \rangle = p^{\vert k\rangle
}_{\alpha\beta} \vert k \,  \rangle \; , \qquad \mathrm{rank}
(p^{\vert k\rangle}_{\alpha \beta}) = n-k \qquad \hbox{((32-k) for
 n=32 $\Leftarrow$ D=11)}\; .
\end{eqnarray}
The definition of BPS preon implies its identification with the state $\vert (n-1)
\rangle$ with  a generalized momentum eigenvalue matrix $p^{\vert (n-1)
\rangle}{}_{\alpha\beta}= p^{(1)}{}_{\alpha\beta}$ of rank equal to one. Indeed, any
matrix of rank one can be expressed by the direct product of two vectors,
\begin{eqnarray}\label{pBPS=ll}
 p_{\alpha\beta}= \lambda_\alpha \, \lambda_\beta \quad \Leftrightarrow \quad
 p_{\alpha\beta}=  p^{\vert (n-1) \rangle}_{\alpha\beta} :=  p^{(1)}_{\alpha\beta} =
\lambda_\alpha \, \lambda_\beta \;
 , \qquad \alpha, \beta =1, \ldots , n \; .
 \end{eqnarray}
 Clearly Eq. (\ref{pBPS=ll}) provides  \cite{BL98}
a generalization of the Cartan--Penrose representation (\ref{P=ll4D}) for the
light--like four vector in the Minkowski spacetime $\mathbf{M}^4$.

In the supersymmetric theory, where $P_{\alpha \beta}= \{ Q_\alpha
\, \, Q_\beta\}$, (see Sec. III) the classification by the rank of
the generalized momentum matrix ($32-k$) provides the
classification of the BPS states by the number of preserved
supersymmetry ($k$) \cite{BPS01}. Here the states {\it  $\vert k
\; \rangle$ are the BPS states preserving $k$ of the $n$
supersymmetries generated by $Q_\alpha$-s}.  The BPS preons
 preserves {\it all but one
supersymmetries}, $\vert BPS\; preon \rangle = \vert (n-1) \, \rangle$, which means
$31$ out of $32$ supersymmetries in the M--theoretic ('$D=11$') case, $\vert BPS\;
preon \rangle = \vert 31 \,  \rangle$.

Now we are ready to discuss the r\^ole of BPS preons as possible
constituents. Notice that a symmetric $n\times n$ matrix always
can be diagonalized by $GL(n)$ transformations, {\it i.e.} there
exists a matrix $ g_{\alpha}{}^{(\gamma)}\in GL(n,\mathbf{R})$
such that
\begin{equation}\label{p=GpG}
p^{\vert k \rangle}_{\alpha\beta} :=  p_{\alpha\beta}^{(32-k)}=
g_{\alpha}{}^{(\gamma)}p_{(\gamma)(\delta)} g_{\beta}{}^{(\delta)}
\end{equation}
 with some diagonal matrix $p_{(\gamma)(\delta)}= diag (\ldots )$ holds.
Moreover, this diagonal matrix can be put in the form $p_{(\gamma)(\delta)}= diag
(1,\ldots,1,-1,\ldots, -1, 0,\ldots,0)$,  where the number of nonvanishing elements,
all $+1$ or $-1$, is equal to $\tilde{n}=(n-k)=rank (p^{(k)}{}_{\alpha\beta})$.

Only at this stage we really need in a reference on supersymmetry.
Indeed, the usual assumptions of unitary supersymmetric quantum
mechanics do not allow for negative eigenvalues of
$P_{\alpha\beta}=\{Q_\alpha , Q_\beta\}$.  Thus, only positive
eigenvalues are allowed and
\begin{equation}\label{pdiag}
p_{(\gamma)(\delta)}= diag
(\underbrace{1,\ldots,1}_{\tilde{n}=32-k},
\underbrace{0,\ldots,0}_{k})\; .
\end{equation}
Substituting (\ref{pdiag}) into (\ref{p=GpG})  and  denoting $
g_{\alpha}{}^{1}=
 \lambda_{\alpha}{}^{1}$, $\ldots$, $ g_{\alpha}{}^{\tilde{n}}=
 \lambda_{\alpha}{}^{\tilde{n}}$, one finds
\begin{eqnarray}\label{npreon}
P_{\alpha\beta}&\vert BPS, k \rangle& = \sum\limits_{r=1}^{\tilde{n}:=32-k}
\lambda_{\alpha}{}^r \lambda_{\beta}{}^r \vert BPS, k \rangle \; \equiv
\left(\lambda_{\alpha}{}^1\lambda_{\beta}{}^1 + \ldots + \lambda_{\alpha}{}^{\tilde{n}}
\lambda_{\beta}{}^{\tilde{n}} \right) \vert BPS, k \rangle \; . \qquad
\end{eqnarray}
Eq. (\ref{npreon}) may be treated  as a manifestation of the {\it composite structure}
of any BPS state $\vert BPS, k \rangle$ with $k<(n-1)$. To this end one  solves
(\ref{npreon}) by
\begin{eqnarray}\label{k=npreon0}
\vert BPS, k \rangle = \vert \lambda^1 \rangle \otimes \ldots
\otimes  \vert \lambda^{(32 - k)} \rangle \; ,
\end{eqnarray}
which implies  that the BPS states  $\vert BPS, k \rangle$ with $k<(n-1)$   are
composed from $\tilde{n}=32-k$ BPS preonic states $ \vert \lambda^1 \rangle$, $\ldots$,
$ \vert \lambda^{\tilde{n}}\rangle$ characterized by the spinors $\lambda_\alpha{}^1$ ,
$\ldots$, $\lambda_\alpha{}^{\tilde{n}}$. Clearly for the vacuum states preserving all
supersymmetries, $k=n$, Eq. (\ref{k=npreon0}) does not make sense; for $k=(n-1)$ it
just identifies different notations for a BPS preon $\vert BPS, 31 \rangle =
 \vert \lambda^1 \rangle$ {\it i.e.} it implies that BPS preons are
 fundamental.

In the light of a supersymmetric treatment this implies that \cite{BPS01} any BPS
states preserving some (but not all)  supersymmetries  can be considered as a composite
of BPS preons. In particular all the M--theory BPS states can be considered as composed
of BPS preons, which allowed us to conjecture that the BPS preons may be considered as
fundamental constituents of M--theory \cite{BPS01}.

The supersymmetry is important in the following respect. {\it In non--supersymmetric}
theory the generalized momentum matrix is not positive definite. This implies the
possibility of minus signs in the diagonalized form of the generalized momentum matrix,
{\it i.e.} $p_{(\gamma)(\delta)}= diag (1,\ldots,1,-1,\ldots, -1, 0,\ldots,0)$ rather
than (\ref{pdiag}). Then, to compose the state with such an eigenvalue of the
generalized momentum one should introduce, in addition to  BPS preons (\ref{(P-ll)pr}),
their counterparts {\it with negative energy}, ''antipreons'' $\vert {anti-BPSpreon}\,
, \, \lambda_\alpha \rangle \equiv \vert {anti-}\lambda_\alpha \rangle $ obeying
$P_{\alpha\beta}\vert {anti-}\lambda_\alpha \rangle = - \lambda_{\alpha}\lambda_{\beta}
\vert {anti-} \lambda_\alpha \rangle$.

\subsection{II.3. BPS preon and generalized Penrose correspondence.
\\ Symplectic  twistors and tensorial spaces}

In the light of the discussion of the first sections, one may
expect that some generalization of the Penrose correspondence
(\ref{m=xl}) should be related with the generalization
(\ref{pBPS=ll}) of the Cartan--Penrose representation
(\ref{P=ll4D}). As (\ref{m=xl}) incudes the coordinate $x^a$
conjugate to the momentum $p_a$ entering in (\ref{P=ll4D}), one
may expect that the desired generalization of the (\ref{m=xl})
should include a spin--tensorial coordinate $X^{\alpha\beta}=
X^{\beta\alpha}$ conjugate to $p_{\alpha\beta}$ of
(\ref{pBPS=ll}). In such a way one arrives at the $n(n+1)/2$
dimensional spacetime with coordinates $X^{\alpha\beta}$,
\begin{eqnarray}\label{Sn0}
& \Sigma^{({n(n+1)\over 2}|0)}\; :  \; X^{\alpha\beta}=
X^{\beta\alpha}\; , \quad
 \alpha, \beta = 1,2, \ldots , n  \;\; ,\quad
\end{eqnarray}
which is called ''tensorial space'' \cite{Fr86,V01s,BPST04}. To
justify this name one can notice that {\it e.g.} for  $n=4$  one
can decompose the symmetric spin--tensorial coordinate
$X^{\alpha\beta}=X^{\beta\alpha}$ of the $\Sigma^{({4(4+1)\over
2}|0)}= \Sigma^{({10}|0)}$ space  on the basis of $D=4$ Dirac
matrices
\begin{eqnarray}
X^{\alpha\beta}&=& X^{\beta\alpha}={1\over
  2}x^\mu\gamma_\mu^{\alpha\beta}+{1\over 4}
y^{\mu\nu}\gamma_{\mu\nu}^{\alpha\beta}\,, \quad \mu\, ,\, \nu=0,1,2,3\,; \quad
\alpha,\beta=1,2,3,4\,, \label{x4}
\end{eqnarray}
arriving at the set of antisymmetric tensorial coordinates
$y^{\mu\nu}=- y^{\nu\mu}$ in addition to the standard four--vector
coordinates $x^\mu$ (see \cite{Fr86}). For $n=16$ one can use the
decomposition on the basis of $D=10$ sigma--matrices,
\begin{eqnarray} \label{x10}
 X^{\alpha\beta}= X^{\beta\alpha} & ={1\over 16}x^\mu\tilde{\sigma}_\mu^{\alpha\beta}+
 {1\over 2\cdot 16 \cdot 5!}  y^{\mu_1\ldots\mu_5}
      \tilde{\sigma}_{\mu_1\ldots\mu_5}^{\alpha\beta}\; , \qquad \mu\,  ,\,  \nu =0,1, \ldots, 9\; ,
      \qquad
 \\ \nonumber &
 \alpha\, ,\beta=1,\ldots ,16\,, \quad y^{\mu_1\ldots\mu_5}= y^{[\mu_1\ldots\mu_5]} =  (-) {1\over 5!}
\epsilon^{\mu_1\ldots\mu_5\nu_1\ldots\nu_5} y_{\nu_1\ldots\nu_5}
\;
\end{eqnarray}
one arrives at the parametrization of $\Sigma^{({10(10+1)\over
2}|0)}= \Sigma^{(55|0)}$ space by $10$ usual vector and $45$
antisymmetric (anti-)selfdual 5--index tensorial coordinates
$y^{\mu_1\ldots\mu_5}= y^{[\mu_1\ldots\mu_5]}= (-) {1\over 5!}
\epsilon^{\mu_1\ldots\mu_5\nu_1\ldots\nu_5} y_{\nu_1\ldots\nu_5}$
(see \cite{EiS}). Finally, in $n=32$ one can use the set of $D=11$
gamma matrices to arrive at the parametrization of
$\Sigma^{(528|0)}$ by the set of vectorial, $x^\mu$, two--index
tensorial $y^{\mu\nu}=- y^{\nu\mu}$ and five--index tensorial $
y^{\mu_1\ldots\mu_5}= y^{[\mu_1\ldots\mu_5]}$ coordinates,
\begin{eqnarray} \label{x11}
 X^{\alpha\beta}= X^{\beta\alpha}& ={1\over 32}x^\mu {\Gamma}_\mu^{\alpha\beta}
 + {i\over 64 \cdot 5!} y^{\mu\nu}
      {\Gamma}_{\mu\nu}^{\alpha\beta}\; + {1\over 32\cdot 5!} y^{\mu_1\ldots\mu_5}
      {\Gamma}_{\mu_1\ldots\mu_5}^{\alpha\beta}\; , \qquad
 \\ \nonumber & \mu\, \nu=0,1, \ldots,10\,; \quad
 \alpha\, ,\beta=1,\ldots ,32\, .  \qquad
\end{eqnarray}
These spaces appear as the bosonic body of the `generalized'
\cite{BL98} or 'extended'/`enlarged' \cite{JdA00,30/32} or
`tensorial' (see \cite{BPST04} and refs. therein) superspaces
$\Sigma^{(10|4)}$, $\Sigma^{(55|16)}$ and $\Sigma^{(528|32)}$ which
we will discuss in Secs III, IV.

The generalized Penrose correspondence  has the simple form \cite{BL98,BPS01}
\begin{eqnarray}
\label{mu00}  \mu^\alpha = X^{\alpha\beta} \lambda_{\beta} \; ,
\qquad \alpha \, , \, \beta \,  = \, 1, \ldots , n \;
\end{eqnarray}
involving the $\Sigma^{(n(n+1)/2|n)}$ coordinates $X^{\alpha\beta}$
and  $2n$ real components ($n$ in $\mu^\alpha$ and $n$ in
 $\lambda_\alpha$) of {\it symplectic twistor}
$\Upsilon_{0\hat{\alpha}}$
\begin{eqnarray}
\label{Up00} \Upsilon_{0\hat{\alpha}}= \left(\mu^\alpha \, , \,
\lambda_\alpha\right)\; .
\end{eqnarray}
This parametrizes the space
 $\mathbf{R}^{2n}-\{ 0\}$ of the fundamental representation
 the $Sp(2n)$ group which leaves invariant the matrix
\begin{eqnarray}\label{C0}
C_{\hat{\alpha}\hat{\beta}}\, = \, \left(\matrix{ 0 &
\delta_{\alpha}{}^{\beta}\cr - \delta^{\alpha}{}_{\beta}&
0}\right)\; \qquad \hat{\alpha}, \hat{\beta} =1 , \ldots , 2n\; ,
\qquad {\alpha}, {\beta} =1 , \ldots , n\; .
\end{eqnarray}
The homogeneity of Eq. (\ref{mu00}) allows one to treat it as an
equation in the {\it projective symplectic twistor space} $
\mathbf{RP}^{2n-1}$.

In distinction to (\ref{m=xl}) {\it with real $x^a$} (and in analogy with
(\ref{m=xLl0}) {\it with complex $x_L^a$}),  Eq. (\ref{mu00}) with an arbitrary real
$X^{\alpha\beta}=X^{\beta\alpha}$ does not set any restrictions on the bosonic spinors
(or s--vectors \cite{V01s})\footnote{One may find better to  call $\lambda_\alpha$ and
$\mu^\alpha$ {\it s-vectors} \cite{V01s} as the invariance of our basic equations is
given by $GL(n)$ and, non-manifestly, by $Sp(2n)$ rather than by some thier $SO(1,D-1)$
subgroup.} $\mu$ and $\lambda$. Indeed, defining $\mu$ and $\lambda$ in an arbitrary
manner one always can find symmetric $X^{\alpha\beta}$ such that (\ref{mu00}) holds.
Moreover, such $X^{\alpha\beta}$ is not unique. The generalized Penrose correspondence
(\ref{mu00}) is  invariant under the $n(n-1)/2$--parametric generalization of the
$b$--symmetry (\ref{bsym-xll}) (see \cite{BL98,B02})
\begin{eqnarray}
\label{vbX00} \delta_{b} X^{\alpha\beta} &=& b^{IJ} u_I^{\alpha}
u_J^{\beta}\;  \quad
 (\qquad  \delta_{b} \lambda_{\alpha}\, =0 \; ,
  \qquad  \delta_{b} \mu^{\alpha}\, =0 \; \qquad ) \; ,
\end{eqnarray}
where $u_I^{\alpha}$,  spinors which are orthogonal to
$\lambda_\alpha$,
\begin{eqnarray}
\label{uI=def0} u^\alpha_I \, \lambda_\alpha =0 \; , \qquad I=1,
\ldots , (n-1) \; .
\end{eqnarray}
Eq. (\ref{vbX00}) provides the general solution of the condition
\begin{eqnarray}
 \label{vbX00=def} \delta_{b} X^{\alpha\beta} \lambda_\alpha = 0 \;
\end{eqnarray}
which can be used as an alternative definition of the generalized $b$--symmetry.
\footnote{To see that the transformations (\ref{vbX00}) provide the counterpart of the
`standard' $b$--symmetry (\ref{b-sym}), one notices that, allowing for the existence of
a non--degenerate antisymmetric matrix $C_{\alpha\beta}= - C_{\beta\alpha}$ [clearly,
for even $n$, see footnote 3; this also reduces $GL(n)$ symmetry down to $Sp(n)$], one
can use its inverse $C^{\alpha\beta}$  to define $\lambda^\alpha = C^{\alpha\beta}
\lambda_{\beta}$ which clearly obeys $\lambda^\alpha \lambda_\alpha=0$. This allows us
to identify this $\lambda^\alpha$ with one of the $u_I^\alpha$ [$u_I^\alpha=
(\lambda^\alpha\, , \, u_{\tilde{I}}^\alpha)$, see \cite{BPS03}] and to find among
(\ref{vbX00}) the counterpart $\delta_b X^{\alpha\beta}= b \lambda^\alpha
\lambda^\beta$  of the $D=4$ transformations (\ref{b-sym}); for $D=3$ there is only one
$u_I^\alpha$ which coincides with $\lambda^\alpha$ and thus all the $b$--symmetry is
reproduced by the above formula.}

Thus Eq. (\ref{mu00}) can be considered as a correspondence between
the space of $(n-1)$ dimensional hyperplanes  in $\mathbf{R}^{2n-1}$
(each parameterized by spinor $\lambda$ modulo its scaling factor
considered to be common with $\mu^\alpha$) and the space of
$n(n-1)/2$ dimensional surfaces (given by
$\hat{X}^{\alpha\beta}(b^{IJ})= {X}^{\alpha\beta} + b^{IJ}
u_I^{\alpha}u_J^{\beta}$) in $n(n+1)/2$ dimensional tensorial space
$\Sigma^{(n(n+1)/2|0)}$.
 On  the language of generalized particle--like
mechanics this correspondence implies that the action
 \cite{BL98}
\begin{eqnarray}
\label{p=0preon0} S_0 =  \int_{W^1} \lambda_\alpha \lambda_\beta
d\hat{X}^{\alpha\beta} \equiv {1\over 2} \int d\tau \,
\lambda_{\alpha}(\tau) \lambda_{\beta}(\tau)
\partial_\tau \hat{X}^{\alpha\beta}(\tau)  \; ,
\end{eqnarray}
 which possess a gauge
$b$--symmetry given by pull--back of Eq. (\ref{vbX00=def}) on the
worldline $W^1$, allows for the reformulation in terms of the
symplectic twistor coordinate functions
$\hat{\Upsilon}_{0\alpha}(\tau)= (\hat{\mu}^\alpha\, , \,
\hat{\lambda}_\alpha )$. Indeed, one can use
 the Leibniz rule ($\hat{\lambda} \partial
\hat{X}\equiv
\partial (\hat{\lambda} \hat{X})-
(\partial \lambda) \hat{X}$) to present the action
(\ref{p=0preon0}) in the equivalent form
\begin{eqnarray}\label{p=0TW0}
S & = & \int_{W^1} \left(d \hat{\mu}^\alpha \,
\hat{\lambda}_{\alpha} - d\hat{\lambda}_{\alpha}\;
\hat{\mu}^\alpha \right) \equiv  \int_{W^1}
d\hat{\Upsilon}_{0\hat{\alpha}}\,
C^{\hat{\alpha}\hat{\beta}}\,\hat{\Upsilon}_{0\hat{\beta}}
  \; ,
\end{eqnarray}
where $\hat{\mu}^\alpha$ is defined by the pull--back of the
generalized Penrose correspondence relation (\ref{mu00}),
$\hat{\Upsilon}_{0\hat{\alpha}}$ by the pull--back of (\ref{Up00})
and $C^{\hat{\alpha}\hat{\beta}}$ is the $Sp(2n)$ invariant of Eq.
(\ref{C0}).

Notice that the action (\ref{p=0preon0}) produces the generalized Cartan--Penrose
representation (\ref{pBPS=ll}) as a primary constraint (see \cite{Dirac}) in the
Hamiltonian formalism. This gives a reason (see the supersymmetric considerations in
\cite{30/32,BPS03} and Sec. III for more) to treat (\ref{p=0preon0}) as a {\it
dynamical model for a point--like BPS preon}.

Hence a search for a generalized Penrose correspondence related to the generalized
Cartan--Penrose representation (\ref{pBPS=ll}) and the definition of the BPS preon
(\ref{(P-ll)pr}) leads us, through Eq. (\ref{mu00}), to the {\it (ortho)symplectic
twistors} (\ref{Up00}) and to the {\it generalized or tensorial (super)spaces}
(\ref{Sn0}) which appeared also in different perspective, see \cite{JdA00,JdA04}.

\subsection{II.4. BPS preons and higher spin fields in $D=4,6,10$}

In distinction to (\ref{Sb4D=mdl+c}), the ''preonic'' action (\ref{p=0TW0}) contains an
{\it unconstrained twistors}, {\it i.e.} no counterpart of the helicity constraint
(\ref{S-U(1)}) appears when one writes the equivalent twistor representation
(\ref{p=0TW0}) of Eq. (\ref{p=0preon0}).  After  quantization of particle mechanics
(\ref{Sb4D=mdl+c}) one arrives at the wave function $\phi(\lambda_A\, ,
\bar{\lambda}_{\dot{A}})$ subject to the quantum counterpart of the constraint
(\ref{lyl=0}). Just the latter constraint makes $\phi(\lambda_\alpha)= \phi(\lambda_A\,
, \bar{\lambda}_{\dot{A}})$ to describe
 massless particle of {\it a certain} helicity (see \cite{Ferber,Balachandran,EiS,B90}, \cite{BLS99}
 and refs. therein), {\it
e.g.} of helicity equal to zero. This explains  the  helicity constraint name. Now, the
action (\ref{p=0TW0}) for the
 $n=4$ case differs from that in Eq.
(\ref{Sb4D=mdl+c}) just by the absence of the helicity constraint (\ref{lyl=0})
($\lambda_{\alpha}=(\bar{\lambda}^{\dot{A}}\, , \, \lambda_A\, )$,
$\mu^{\alpha}=(\bar{\mu}_{\dot{A}}\, , \, \mu^A\, )$). Hence, one may expect that the
quantum state spectrum of this ``preonic'' particle mechanics would include a tower of
massless field of all possible helicities. The analysis of \cite{BLS99} showed that
this is indeed the case  for $n=4$ $D=4$ and indicated
 an infinite tower of massless $D=6$ and $D=10$ `higher spin'
fields in the model with $n=8$ and $n=16$. As it is finally shown in \cite{BBdAST04}
these are all the {\it conformal massless} higher spin fields in $D=6$ and $D=10$
dimensions, respectively.

Here we will try to create some image of the relation between preonic particle
mechanics and higher spin fields with an emphasis on the r\^ole of twistor--like
methods and notions. More technical details can be found in the original papers.

\medskip

\centerline{\it II.4.1. Higher spins from tensorial space.}
\medskip
Interestingly enough, the tensorial space (\ref{Sn0}) with $n=4$, Eq. (\ref{x4}), was
proposed in \cite{Fr86} as a basis for the construction of D$=$4 higher--spin theories.
It was known that a consistent interaction of higher spin fields requires i) {\it an
infinite tower of all possible higher spin fields} and ii) a spacetime with a nonzero
cosmological constant (see \cite{Vasiliev89,Misha0304b,Dima04}). The assumption of
\cite{Fr86}  was that there may exist a theory in a ten--dimensional space
$\Sigma^{(10|0)}$ whose (alternative--to--) Kaluza--Klein reduction may lead in $D=4$
to an infinite tower of fields with increasing spins instead of the infinite tower of
Kaluza--Klein particles of increasing mass. It was argued that the symmetry group of
the theory should be $Sp(8)\supset SU(2,2)$, and $OSp(1|8)$ in supersymmetric case. The
idea was that using a single representation of $OSp(1|8)$ (such that it contains each
and every massless higher spin representation of the $D=4$ superconformal group
$SU(2,2)\subset OSp(1|8)$ only once) in the ten--dimensional tensorial space one could
describe an infinite tower of massless higher spin fields in $D=4$ space-time.

In this perspective the $\Sigma^{(n(n+1)/2|n)}$ superparticle action of \cite{BL98},
the {\it point--like model for BPS preon} in the light of \cite{BPS01} and
\cite{B02,30/32,BPS03}, provided (rather accidentally; in its $n=4$ $\Sigma^{(10|4)}$
version) a dynamical realization of the proposal from \cite{Fr86}. This {\it preonic
superparticle action}, whose purely bosonic limit  is given by Eq. (\ref{p=0preon0}),
involves the auxiliary bosonic spinor variables $\lambda_\alpha(\tau)$. These provide a
{\it twistorial dimensional reduction} (for $n>2$) and, for $n=4,8,10$, also a {\it
twistorial compactification} mechanism \cite{BLS99} which  results in the discreetness
of the quantum state spectrum and its identification, for $n=4$, with the spectrum of
all the massless higher spin fields in $D=4$ \cite{BLS99} and, for $n=8,16$, with the
spectrum of all {\it conformal} massless higher spin fields in $D=6$ and $10$
\cite{BBdAST04}. The AdS generalization of the model \cite{BL98} is provided by the
superparticle on the $OSp(1|n)$ supergroup manifold \cite{BLS99,BLPS}. It was
conjectured in \cite{BLPS,V01s} and shown in \cite{Misha+,Dima} that a field theory on
$OSp(1|4)$ is classically equivalent to the $OSp(1|8)$--invariant free higher spin
field theory in $AdS_4$.

The preonic particle  model (\ref{p=0preon0}) possesses manifest
$GL(n)$ and non--manifest  $Sp(2n)$ symmetry ($OSp(1|2n)$ in
supersymmetric case), thus showing the expected symmetry of higher
spin theories. The latter becomes manifest symmetry after passing
to an equivalent twistor form (\ref{p=0TW0}). The $Sp(2n)$
symmetry is also manifest in the following equivalent form  of the
action (\ref{p=0preon0}) \cite{V01s}
\begin{eqnarray}
\label{p=0pr-mu} S_0 =  \int_{W^1} \left( \lambda_\alpha
\lambda_\beta d\hat{X}^{\alpha\beta} +  \tilde{\mu}^\alpha\,
d\lambda_\alpha \right) \; .
\end{eqnarray}
One of the simple ways to show the equivalence of
(\ref{p=0preon0})  and (\ref{p=0pr-mu})  is to  notice
 that, moving the derivatives, one can rewrite
 (\ref{p=0pr-mu}) in the equivalent form  (\ref{p=0TW0}) of the action
(\ref{p=0preon0}). The only difference then will be
  a shift in
 definition of  $\mu^\alpha$: Eq. (\ref{mu00}) is replaced by
 $\; \mu^\alpha = X^{\alpha\beta}\lambda_\beta +
 \tilde{\mu}^\alpha$.

For Hamiltonian formalism the use of the action (\ref{p=0pr-mu}) instead of
(\ref{p=0preon0}) looks like  a simple method of {\it conversion} of the second class
constraints, which are present for (\ref{p=0preon0}), into the first class ones (see
\cite{Dima03} for further discussion and references; the conversion was also done in
\cite{BLS99} but in a more complicated way). The quantization with such a conversion
results in the preonic wave function $\Phi(X, \lambda)$ subject to the constraint
\cite{BLS99}
\begin{eqnarray}
\label{(P-ll)Phi(X,l)} \left( \partial_{\alpha\beta} - i
\lambda_{\alpha}\lambda_{\beta} \right) \, \Phi(X, \lambda) \; = 0
\; , \qquad  \partial_{\alpha\beta}:= {\partial /
\partial {X}^{\alpha\beta} } \; .
\end{eqnarray}
Eq. (\ref{(P-ll)Phi(X,l)}) is clearly the coordinate
representation of the definition of the BPS preon (\ref{(P-ll)pr})
{\it provided the preon is considered as a point--like} object in
tensorial space,
\begin{eqnarray}
 \label{Phi(X,l)}
 \Phi(X, \lambda) = < X^{\alpha\beta} \, \vert \; \lambda_\alpha
 \rangle \equiv < X^{\alpha\beta} \, \vert \; BPS\; preon\; \lambda_\alpha
 \rangle \; .
\end{eqnarray}
This gives one more reason to state that the  generalized (super)particle model
\cite{BL98} with the bosonic limit (\ref{p=0preon0}) provides a model for a point--like
BPS preon \cite{BPS01,BPS03}.

The solution of  Eq. (\ref{(P-ll)Phi(X,l)}) is given by the
generalized plane wave,  {\it preonic plane wave},
\begin{eqnarray}
\label{prPhi(X,l)} \Phi(X, \lambda) = \phi(\lambda) \, \exp \{ -i
\lambda_\alpha \, X^{\alpha\beta} \, \lambda_\beta\} \; .  \;
\end{eqnarray}
involving an arbitrary function  $\phi(\lambda) $ of the bosonic spinor
$\lambda_\alpha$.\footnote{\label{F6} The choice of the class of functions where
$\phi(\lambda) $ takes its values is not unique. One should fix it to provide the
convergence of integrals used on the way to a spacetime treatment which, in its turn,
is also not unique. This interesting issue is beyond the score of that contribution;
see \cite{V01c,Dima03} and also sec. IIA and footnote 5 in \cite{BLS99} for
discussions. } Its integration over $\lambda$
\begin{eqnarray}
\label{iP(X,l)} b(X)=  \int \, d^n \lambda\;  \Phi(X, \lambda) =
\int \, d^n \lambda\;  \phi(\lambda) \, \exp \{ -i \lambda_\alpha \,
X^{\alpha\beta} \, \lambda_\beta\} \;  \;
\end{eqnarray}
provides the general solution of the following equations in
tensorial space
\begin{eqnarray}
\label{ddb=0}
\partial_{\alpha [\beta } \partial_{\gamma ]\delta }
 b(X)\equiv 1/2 (\partial_{\alpha \beta } \partial_{\gamma \delta }
 - \partial_{\alpha \gamma } \partial_{\beta\delta })
 b(X)= 0 \; .
\end{eqnarray}
This was proposed in \cite{V01s} as  dynamical equations for
massless higher spin fields. It was shown in \cite{V01s} (see also
\cite{Dima03}) that, for $n=4$, decomposing the field $b(X)=b(x^\mu
\, , \, y^{\mu\nu})$ (see (\ref{x4})) in the series on $y^{\mu\nu}=
-y^{\nu\mu}$ one finds all the field strengths of the massless {\it
bosonic} $D=4$ higher spin fields {\it i.e.} of the higher spin
fields with integer spin; their equations of motion follow from
(\ref{ddb=0}). The details and further discussion on relation
between field theories in tensorial space and in the standard
spacetime can be found in \cite{V01c,Dima03} (mainly for $D=4$) and
in \cite{BBdAST04} (also for $D=6,10$).

The field strengths of massless fields with all the possible
half--integer spins are collected in the spinorial field
$f_\alpha(X)$ obeying a tensorial space counterpart of the Dirac
equation
\begin{eqnarray}
\label{df=0}
\partial_{\alpha [\beta } f_{\gamma ]}(X)\equiv
1/2 (\partial_{\alpha \beta } f_{\gamma }(X) - \partial_{\alpha
\gamma}f_{\beta}(X)) = 0 \; .
\end{eqnarray}
The general solution of Eq. (\ref{df=0}) is given by the integral of
the preonic plane wave (\ref{prPhi(X,l)}) with measure $d^n\lambda\,
\lambda_\alpha$,
\begin{eqnarray}
\label{ifP(X,l)} f_\alpha (X)=  \int \, d^n \lambda\; \lambda_
\alpha \, \Phi(X, \lambda) = \int \, d^n \lambda\;  \lambda_
\alpha \, \phi(\lambda) \, \exp \{ -i \lambda_\alpha \,
X^{\alpha\beta} \, \lambda_\beta\} \; . \;
\end{eqnarray}
Clearly, as for the even $n$ (including $n=4$) the measure is
symmetric under $\lambda \mapsto - \lambda$, the half integer
fields collected in $f_\alpha (X)$ come from the odd part of
$\phi(\lambda)$, $\phi_-(\lambda)=1/2 (\phi(\lambda) -
\phi(-\lambda))$, while the integer fields collected in $b(X)$
come from the even part of $\phi(\lambda)$, $\phi_+(\lambda)=1/2
(\phi(\lambda) + \phi(-\lambda))$.  However both integer and
half--integer fields come from the same ''twistorial wave
function'' $\phi(\lambda)$.
 \footnote{\label{7} In this respect one notices (see
\cite{BLS99} and \cite{BBdAST04}) that the quantum state spectrum of
the preonic particle is already supersymmetric (contains all integer
and all half integer fields) while the spectrum of its
supersymmetric generalization  \cite{BLS99} is doubly degenerate. To
resolve the degeneracy and to provide the physical spin--statistics
correspondence one uses a projection relating the Grassmann parity
and the parity with respect to $\lambda\mapsto - \lambda$; see
\cite{BLS99}, \cite{BPST04} and refs. therein for further
discussion.} It appears directly as a result of quantization when
one starts from the action (\ref{p=0TW0}).

\medskip \centerline{\it II.4.2. Twistor wave function,
Cartan--Penrose representation and the Hopf fibrations.}
\centerline{ D=3,4,6 and 10.}
\medskip
A simplest way to quantize the preonic particle model
\cite{BL98} is to use an equivalent twistor representation
(\ref{p=0TW0}) of the preonic action (\ref{p=0preon0}) \cite{BLS99}.
Indeed,  the action (\ref{p=0TW0}) is i) written in terms of
unconstrained symplectic twistor $(\mu^\alpha\, , \, \lambda_\alpha
)$; ii) is the first order action, which allows to identify
$(\mu^\alpha$ with the momentum conjugate to the coordinate
$\lambda_\alpha$ (or {\it vice versa}); iii) the Hamiltonian of the
system is identically zero, which implies that the system is free
and that, after quantization, the Schr\"odinger equation just states
an independence of the proper time parameter $\tau$
(reparametrization invariance). As a result, one sees that the wave
function of the preonic particle is an arbitrary function
$\phi(\lambda_\alpha)$ of the bosonic spinor $\lambda_\alpha$ (see
footnote \ref{F6}).

To provide the spacetime treatment of such a wavefunction one
applies \cite{BLS99} the generalized Penrose correspondence
(\ref{pBPS=ll}) and extracts the $D$ dimensional momentum from the
generalized momentum using the $n\times n$ real (or pseudo--real)
representation of $D$--dimensional gamma matrices ($n=2^{[D/2]}$ for
$D\not=10$  and $n=2^{[D/2]}$ for other $D$)
\begin{eqnarray}\label{Pnn=ll}
p_{\alpha\beta}= \lambda_{\alpha}\lambda_{\beta} \; , \qquad  p_\mu
 \equiv \, {1\over n} \, p_{\alpha\beta} \Gamma_\mu^{\alpha\beta} \, = \,
 \lambda \Gamma_\mu \lambda \, \qquad
\cases{\alpha, \beta = 1, \ldots , n \cr \mu= 0, 1, \ldots , D-1}
\end{eqnarray}
Thus the wave function $\phi(\lambda)$ can be treated as dependent
on the spacetime momentum $p_\mu$ and the additional variables
parameterizing the fibration  $\Im^D_n$  of the space $\pounds :=
\mathbf{R}^n-\{ 0\}= \mathbf{S}^{n-1}\otimes \mathbf{R}_+$ of
nonvanishing spinors $\lambda_\alpha \not= (0, \ldots , 0)$ over the
(base) space $\wp := \{ p_\mu \, : p_\mu = \lambda \Gamma_\mu
\lambda \}$ of momentums  determined by the Cartan--Penrose
representation $p_\mu = \lambda \Gamma_\mu \lambda$,
\begin{eqnarray}\label{ph(l)=} \phi(\lambda)= \phi(p\, ;\,
\Im^D_n)\vert_{p=\lambda\Gamma\lambda}\quad  \Leftrightarrow
\qquad \phi(\pounds) = \phi(\wp\, , \, \Im)\, , \quad \Im \approx
&  { \pounds \over \wp}
\end{eqnarray}
 The properties of the
base space $\wp= \{ p_\mu \, : p_\mu = \lambda \Gamma_\mu \lambda
\}$ depends strongly on $D$ and $n$.

For $n=2,4,8,16$ corresponding to $D=3,4,6,10$  the famous identity
$\Gamma^{\mu(\alpha\beta} \Gamma_\mu{}^{\gamma)\delta}\equiv 0$
holds. It results in a light--like momentum $p_\mu$
\begin{eqnarray}\label{Pll2=0}
 D=3, 4, 6, 10 \; : \quad p_\mu
 = \, \lambda \Gamma_\mu \lambda \, \quad \Rightarrow \; p_\mu p^\mu
 = 0 \; , \quad
\end{eqnarray}
Hence the space $\wp$ spanned by momenta $p_\mu= \lambda \Gamma_\mu
\lambda $ is $D-1$ dimensional (rather than $D$--dimensional),
$\wp= \mathbf{R}^{(D-1)}-\{ 0\}= \mathbf{S}^{(D-2)}\otimes
\mathbf{R}_+$. Hence the space $\Im_D^n$ of additional variables in
(\ref{ph(l)=}) is the fibration $\pounds^n/\wp$ of $\pounds=
\mathbf{S}^{(n-1)}\otimes \mathbf{R}_+$ over $\wp=
\mathbf{S}^{(D-2)}\otimes \mathbf{R}_+$ which, in the light of
identification of scales by $p_\mu= \lambda \Gamma_\mu \lambda $,
 is the fibrations of spheres over spheres, $
\Im^n_D := {\pounds^n \over \wp^D}=
 {\mathbf{S}^{(n-1)}\otimes
\mathbf{R}_+\over \mathbf{S}^{(D-2)}\otimes \mathbf{R}_+}
={\mathbf{S}^{(n-1)}\over \mathbf{S}^{(D-2)}}$. Furthermore, as in
the dimensions $D=3,4,6$ and $10$ the number of components of
minimal real (or, in $D=6$, pseudoreal) representation can be
written as $n=2(D-2)$ (see \cite{Dima} and refs. therein), the
spaces of nonvanishing bosonic spinors  are $\pounds^n=
\pounds^{2(D-2)}= \mathbf{S}^{(n-1)}\otimes \mathbf{R}_+=
\mathbf{S}^{(2D-5)} \otimes \mathbf{R}_+ $ and the fibrations
parametrized by additional variables can be presented in a more
transparent form
 $\Im_D^n =
\Im_D^{2(D-2)}=\mathbf{S}^{(2D-5)}/\mathbf{S}^{(D-2)}$. This form
makes  evident that  for  $D=3, 4, 6$ and $10$ the spaces $\Im$ of
auxiliary variables in (\ref{ph(l)=}) are given by the Hopf
fibration of spheres over spheres which are isomorphic to spheres
$S^{(D-3)}$,
\begin{equation}\label{Hopff}
D=3,4,6,10 \quad : \qquad \Im_D^n \equiv {\mathbf{S}^{(n-1)}\over
\mathbf{S}^{(D-2)}}={\mathbf{S}^{(2D-5)} \over
\mathbf{S}^{(D-2)}}= \mathbf{S}^{(D-3)} \quad \left(=
(\mathbf{Z}_2, \mathbf{S}^{1}, \mathbf{S}^{3},
\mathbf{S}^{7})\right)\; .
\end{equation}

For $D=3$ Eq. (\ref{Hopff}) gives  $\Im_3^2= \mathbf{Z}_2$. Thus the
wave function $\phi(\lambda)$, Eq. (\ref{ph(l)=}), can be treated as
function of a light--like momentum (\ref{Pll2=0}) and a sign
variable $\phi(\lambda)= \phi(p_\mu : p^2=0\, ; \, \pm )$. In the
cases of $D=4,6,10$ the wave function $\phi(\lambda)$ of Eq.
(\ref{ph(l)=}) depends, in addition to the light--like momentum
$p_\mu= \lambda\Gamma_\mu\lambda$, on a number of angular variables
which parameterize the compact spaces $S^{(D-3)}$
\begin{eqnarray}
\phi(\lambda_\alpha) = \Phi(p_\mu \; , \; S^{(D-3)})\vert_{p_\mu
p^\mu=0} & \equiv \Phi(p_\mu \, ,  \alpha_1 \, , \ldots  ,
\alpha_{D-3} )\vert_{p_\mu p^\mu=0\;}  \, , \;  D=4, 6, 10 \, . \;
\end{eqnarray}
Thus {\it for D=4,6,10 the space of additional variable
$\Im_D^{2(D-2)}$ in (\ref{ph(l)=}) is compact and isomorphic to the
spheres $S^{(D-3)}$}. This phenomenon was called {\it twistor
compactification} in \cite{BLS99}. This twistorial compactification
is alternative to the Kaluza--Klein one in particular as it occurs
in momentum space and hence makes  the coordinates discrete. These
discrete ''coordinates'' can be identified with  quantum numbers
 enumerating the possible helicity states of all massless
higher spin fields in $D=4$ \cite{BLS99} and all {\it conformal}
massless higher spin fields in $D=6,10$ \cite{BBdAST04}.

In the quantization of preonic mechanics \cite{BLS99} the twistor
compactification  occurs due to the treatment of the gamma--trace of
generalized momentum $p_{\alpha\beta}$ as spacetime momentum (see
(\ref{Pnn=ll})) and due to the generalized Cartan--Penrose
representation for $p_{\alpha\beta}$. It was also noticed
\cite{BLS99} that the generalized Cartan--Penrose representation
(\ref{Pnn=ll}), $p_{\alpha\beta}= \lambda_{\alpha}\lambda_{\beta}$
provides, for $n>2$,  a mechanism of {\it twistorial dimensional
reduction} which also occur in momentum space and reduces the number
$n(n+1)/2$ of degrees of freedom in $p_{\alpha\beta}$ to the smaller
number $n$ of degrees of freedom in the bosonic spinor
$\lambda_\alpha\;$.

\medskip \centerline{\it II.4.3. Problems in M--theoretical $D=11$
case $n=32$} \medskip What turns out  to be different in the
M--theoretical $D=11$ case is that the momentum $p_\mu= 1/32\,
\lambda \Gamma_\mu \lambda$ {\it is not} light--like, $p_\mu
p^\mu\not= 0$. Its square (the $D=11$ mass operator) can be rather
expressed through the values of tensorial charges, $\; p_\mu p^\mu =
- 2 Z^{\mu\nu}Z_{\mu\nu} - 5! Z_{\mu_1\ldots \mu_5 }Z^{\mu_1\ldots
\mu_5 }$ \cite{BL98-2}, also constructed from the bosonic spinor:
$Z_{\mu\nu} = {i\over 64} \lambda \Gamma_{\mu\nu} \lambda$ and
$Z_{\mu_1\ldots \mu_5 } = {1\over 32\cdot 5!} \lambda
\Gamma_{\mu_1\ldots \mu_5 } \lambda$. Thus, if one identifies
$p_\mu= 1/32\, \lambda \Gamma_\mu \lambda$ with the
eleven--dimensional momentum, this is not restricted by a mass shell
condition and parametrizes $\mathbf{R}^{D}-\{ 0\} =
\mathbf{S}^{(D-1)}\times \mathbf{R}_+= \mathbf{S}^{10}\times
\mathbf{R}_+$ (instead  of $\mathbf{S}^{(D-2)}\times \mathbf{R}_+=
\mathbf{S}^{9}\times \mathbf{R}_+$ as it would be if the momentum
were light--like). Then the additional variables in (\ref{ph(l)=})
parametrize the fibration ${\mathbf{S}^{31}\over \mathbf{S}^{10}}$.
Such a $21$ dimensional space is not (is not known to be) isomorphic
to a sphere or a well--studied manifold. The indefiniteness of the
$p_\mu p^\mu$ for $p_\mu= 1/32\, \lambda \Gamma_\mu \lambda$ in
$D=11$ can be treated as the continuity of the mass spectrum (see
\cite{BL98-2}), which is another possibility left by conformal
invariance of the  particle or particle--like mechanics (the latter
case includes null--(super)--$p$--branes, see \cite{BZ} and refs.
therein).

Notice that a similar problem appeared for eleven---dimensional
supermembrane (now M2--brane) \cite{Hoppe} and that now this is
treated  \cite{Nicolai} as an indication that the supermembrane is a
composite state, a system of ten--dimensional $D$--branes, in the
spirit of Matrix model  \cite{(M)atrix}. One might try to understand
the relation between the D=11 BPS preons and spacetime fields
(''higher spin'' $D=11$ or $D=10$ fields) in such a perspective.
However, for a moment we do not have much to say on this direction.

\section{III. BPS preons and supersymmetry}

We discussed the definition and some properties of BPS preons,
including their relation with massless higher spin theories in
$D=4,6,10$ (almost) without the use of supersymmetry. Now we move to
the supersymmetric aspects of the BPS preon conjecture. The pure
bosonic definition of the BPS preon (\ref{(P-ll)pr}) uses the
generalized momentum. As we have noticed, this appears to be related
with the most general supersymmetry algebra.

\subsection{III.1. Generalized momentum,
M--algebra and BPS states}

For any $n=2^m$ the generalized momentum operator
$P_{\alpha\beta}=P_{\beta\alpha}$ is associated with the bosonic
central generator of most general supersymmetry algebra
characterized by the most general form of the commutator of two
fermionic supercharges  $Q_\alpha$,
\begin{eqnarray}\label{QQP}
{}\{ Q_\alpha , Q_\beta\} &=& P_{\alpha \beta} \; , \quad P_{\alpha
\beta}= P_{\beta\alpha} \; ,  \quad \alpha = 1,2, \ldots n\; ;
\end{eqnarray}
the property of $P_{\alpha\beta}$ to be central is expressed by
\begin{eqnarray}
  {} [Q_\alpha , P_{\beta\gamma}]=0 \; ,\qquad
\label{PP=0}{} && {} [P_{\alpha \beta}, P_{\gamma\delta}]=0 \; .
\quad
\end{eqnarray}
The algebra (\ref{QQP}) with $n=32$ ($\alpha=1,2,\ldots 32$)
\cite{HP82,M-alg} is usually called M--theory superalgebra or
M--algebra \cite{M-alg} \footnote{\label{6}See \cite{JdA00} for a
treatment of (\ref{QQP})--(\ref{PP=0}) as central extension of the
abelian fermionic translation algebra, \cite{gM-alg,JdA00} and refs.
therein for further generalizations of the M--theory superalgebra
and for their structure. For $n\not=2^l$, including odd values of
$n$ one can treat the algebra as (\ref{QQP}) as $d=2$ extended
supersymmetry algebra with central charges. }. It encodes a full
information about the nonperturbative BPS states of the hypothetical
underlying M--theory and also its duality symmetries
\cite{Duality,M-alg}. Indeed, for instance, treating $\alpha, \beta$
as eleven dimensional ($SO(1,10)$) spinor indices, one may decompose
the symmetric spin--tensor generator $P_{\alpha\beta}$ on the basis
provided by antisymmetric products of $D=11$ gamma--matrices
\begin{eqnarray}
& \alpha , \beta  = 1,2, \ldots 32 \quad , \label{n32} & \qquad
P_{\alpha\beta} = P_\mu \Gamma^\mu_{\alpha\beta} + Z_{\mu\nu} \,
i\Gamma^{\mu\nu}_{\alpha\beta} +  Z_{\mu_1\ldots \mu_5}
\Gamma^{\mu_1\ldots \mu_5}_{\alpha\beta}\; .
\end{eqnarray}
Here $P_\mu$ is treated as the $D=11$ spacetime momentum, while the additional central
bosonic generators $Z_{\mu\nu} = - Z_{\nu\mu}=Z_{[\mu\nu]}$, $Z_{\mu_1\ldots
\mu_5}=Z_{[\mu_1\ldots \mu_5]}$ can be treated \cite{JdAT} as topological charges of
the BPS states corresponding to supersymmetric extended objects, branes, living in the
eleven--dimensional world. For instance, the spacial components $Z_{IJ}$ and
$Z_{J_1\ldots J_5}$ of $Z_{\mu\nu}$ and $Z_{\mu_1\ldots \mu_5}$ are related to the
topological charges \cite{JdAT} of the supermembrane  and the super--M5--brane
\cite{JdAT,ST97}.

A simple but important observation is that, {\it e.g.} for $n=32$, the M--algebra
(\ref{QQP}), (\ref{PP=0}) possesses the $GL(32)$ ($GL(n)$) automorphism symmetry, which
is broken down to $SO(1,10)$ only upon the use of decomposition (\ref{n32}). This
implies the ``brane rotating'' nature of the $GL(32)/SO(1,10)$ symmetry \cite{West}. On
the other hand it indicates that the above eleven--dimensional treatment based on the
decomposition (\ref{n32}) is, certainly, not a unique one. Treating $\alpha, \beta$ in
(\ref{QQP}), (\ref{PP=0})  as multiindices and using the set of gamma--matrices for
other $D$ in direct product with the internal space gamma--matrices one may provide the
D=10 type IIA, D=10 type IIB, D=4 N=8 \cite{M-alg}, as well as a more exotic D=2+10
treatment \cite{Bars}. As a result, the information about nonperturbative BPS states
of, say, $D=10$ superstring theories (including Dirichlet superbranes) can also be
extracted from (\ref{QQP}). This also explains why  the M--algebra (\ref{QQP}) encodes
as well all the duality relations between different $D=10$ and $D=11$ superbranes.

\subsection{III.2. BPS preons states preserving all but one
supersymmetry. A classification of M-theory BPS states}

Associating the generalized momentum matrix with the right hand side
of the general supersymmetry algebra (\ref{QQP}) one finds (see
\cite{BPS01,BPS03,30/32}) that the definition of the BPS preon in
Eqs. (\ref{pr=lb}), (\ref{(P-ll)pr}) is equivalent to defining the
BPS preon as a state
 preserving  all supersymmetries but one (hence the
notation $\vert BPS , (n-1) \rangle$; $\vert BPS , 31 \rangle$ in
$D=11$) \cite{BPS01}. The bosonic spinor parameters
$\in_I{}^\alpha$ corresponding to the supersymmetries preserved by
a BPS preon $| \lambda  \rangle$,
\begin{eqnarray}\label{preonSUSY}
\in_I^\alpha  Q_\alpha \vert \lambda \rangle = 0 \;  , \qquad
I=1,\ldots, (n-1) \qquad ( I=1,\ldots,  31 \;\; in\;\; D=11)
\end{eqnarray}
are `orthogonal' to the bosonic spinor $\lambda_\alpha$ that labels
it,
\begin{eqnarray}\label{eIl=0}
 \in_I{}^\alpha \lambda_\alpha =0 \; ,
 \qquad I=1,\ldots, (n-1) \; \qquad ( I=1,\ldots,  31 \;\; in\;\; D=11).
\end{eqnarray}
Notice that these are the same bosonic spinors  as in Eq.
(\ref{uI=def0}), which completes the definition (\ref{vbX00}) of the
$b$--symmetry transformations,
 \begin{eqnarray}\label{e=u}
 \in_I{}^\alpha =  u_I{}^\alpha \; . \qquad \end{eqnarray}

In general, the number $n-k$ ($32-k$  in D=11) of supersymmetries preserved by a BPS
state $\vert BPS, k \rangle$ coincides \cite{BPS01,30/32,BPS03} with the rank of the
eigenvalue matrix $p^{(k)}{}_{\alpha\beta}$ (\ref{kBPSp0}) of the generalized momentum
$P_{\alpha\beta}$
\begin{eqnarray}\label{kBPSQ}
\in_{\check I}{}^\alpha Q_\alpha \vert BPS\; k \rangle =0 \; ,
\quad {\check I}= 1, \ldots , k \quad \Rightarrow \quad & \cases{
P_{\alpha \beta} \vert BPS\; k \rangle = p^{\vert k\rangle
}_{\alpha\beta} \vert BPS\; k \rangle\; ,\cr \mathrm{rank}
(p^{\vert k\rangle}_{\alpha \beta}) = 32- k \; .}
\end{eqnarray}
In this respect {\it all  the BPS states} related to a general
supersymmetry algebra (\ref{QQP}), including the M-theory BPS states
for $n=32$, {\it may be  classified by the number of preserved
supersymmetries} \cite{BPS01}. This is the same as the
classification by rank of the generalized momentum matrix considered
in Sec. II.1. Then the discussion of section II.2. implies that {\it
a BPS state preserving $k$ supersymmetry can be treated as a
composite  of $\tilde{n}=n-k$ BPS preons}
\begin{eqnarray}\label{k=npreon}
 \in_{\check I}{}^\alpha
Q_\alpha \vert BPS\; k \rangle =0 \; , \quad {\check I}= 1, \ldots ,
k \; \quad &\Longrightarrow & \quad  \vert BPS, k \rangle = \vert
\lambda^1 \rangle \otimes \ldots \otimes  \vert \lambda^{(32 - k)}
\rangle                 \;
  , \qquad \\ \nonumber
\in_{\check I}{}^\alpha \,\lambda^{(1)}_\alpha &= 0& \; ,  \;
\ldots \; , \;\; \in_{\check I}{}^\alpha \,\lambda^{(n -
k)}_\alpha = 0 \; .
\end{eqnarray}

\subsection{III.3. On ``AdS generalizations'' of the M--algebra \\ and
of the BPS preon definition }

Eqs. (\ref{QQP})--(\ref{PP=0}) give the generalization of the
super--Poincar\'e algebra. The corresponding generalization of the
superconformal algebra is suggested to be $OSp(1|2n)$ [see
\cite{DA+F,HP82} as well as  \cite{BL98} and \cite{30/32} for more
references]. One can ask how the analogous generalization of the
AdS superalgebra look like. The study  of
\cite{BLS99,V01s,Misha+,Dima03} suggests that this is given by the
Lie superalgebra of the $OSp(1|n)$ supergroup,
\begin{eqnarray}\label{OSpbb}
{} [P_{\alpha \beta}, P_{\gamma\delta}]= i \, \varsigma \;
(C_{\alpha(\gamma}P_{\delta ) \beta} + C_{\beta(\gamma}P_{\delta )
\alpha}) \; , \\ \label{OSpbf} {} [ P_{\alpha\beta}\, , \,
Q_{\gamma}]= i \, \varsigma\;   C_{\gamma (\alpha} Q_{\beta)}
\\
\label{OSpff} {}\{ Q_\alpha , Q_\beta\} &=& P_{\alpha \beta} \; ,
\end{eqnarray}
where the dimensional parameter $\varsigma$, the inverse AdS
radius, is introduced to make transparent the contraction of
$OSp(1|n)$ down to the generalized superPoincar\'e supergroup
$\Sigma^{(n(n+1)/2|n)}$ with the algebra
(\ref{QQP})--(\ref{PP=0}); this occurs in $\varsigma \mapsto 0$
limit.

Clearly, the noncommutative $P_{\alpha \beta}$ cannot be
diagonalized and the above definition of the BPS preon should be
modified for that case. The study of \cite{Misha+,Dima03} suggests
the following definition of the BPS preons for that case ({\it cf.}
\cite{BPST04})
\begin{eqnarray}\label{PreonOSp}
(P_{\alpha \beta} - Y_{(\alpha} Y_{\beta)}) \vert BPS\;preon\; ;
\, \lambda_{\alpha} \rangle = 0 \; , \qquad Y_{\alpha}:=
\lambda_{\alpha} - \varsigma  P_{\alpha}^{(\lambda)} \; .
\end{eqnarray}
Here $P_{\alpha}^{(\lambda)}$ is the operator of momentum
conjugate to $\lambda_{\alpha}$ and, hence, the spinorial
operators $Y_{\alpha}$ do not commute for  $\varsigma \not= 0$,
\begin{eqnarray}\label{YYOSp}
 {} [ Y_{\alpha}\, , \,  Y_{\beta}] = 2i \varsigma C_{\alpha\beta} \; .
\end{eqnarray}
Thus, in distinction to generalized super--Poicar\'e ($\varsigma
=0$) case, the definition of BPS preons for the generalized AdS or
$OSp(1|n)$ superalgebra refers to a factorization of the
noncommutative spin--tensorial operator rather than to an eigenvalue
problem.

Here, however, we mainly consider the case of the generalized
super--Poincar\'e algebra with central $P_{\alpha\beta}$ which
allows for the above  simple definition (\ref{(P-ll)pr}) of the BPS
preon \cite{BPS01}.

\subsection{III.4. BPS preons and BPS states in supergravity}

As discussed above the BPS states $\vert BPS\; k \rangle$,
$k\not=0$,  preserve a fraction  ($k/n$) of the supersymmetries; due
to this fact they saturate the Bogomolny--Parasad--Sommerfield or
BPS bound (hence the ''BPS state'' name) and, as a result, are
stable.

In supergravity the algebraic notion of BPS state is realized as a
supersymmetric solution of the supergravity equations {\it i.e.} the
solitonic solution  preserving a fraction  $k/n$ ($k/32$ in the
M--theoretic $n=32$ case) of the {\it local supersymmetries}
characteristic of the supergravity theory. The $k$ supersymmetries
(\ref{kBPSQ}) preserved by the BPS state $ \vert BPS\; k \rangle $
are represented in this ''solitonic'' picture by a set of $k$
linearly independent {\it Killing spinors} $\in_{\check J}{}^\alpha
(x)$ obeying the
 {\it Killing spinors equation}
\begin{eqnarray}\label{Killing}
 {\cal D} \in_{\check J}{}^\alpha :=d  \in_{\check J}{}^\alpha -
 \in_{\check J}{}^\beta \omega_\beta{}^\alpha \equiv D\in_{\check J}{}^\alpha
  - \in_{\check J}{}^\beta \omega_\beta{}^\alpha  =0 \; , \qquad
  {\check J}= 1, \ldots , k \; . \;
\end{eqnarray}
In many cases, including  higher dimensional supergravity ({\it
e.g.} $D=10, 11$) and extended supergravity  in $D=4$  ({\it e.g.}
 $N=4,8$) the generalized covariant derivative ${\cal D}$ is
constructed with the use of {\it generalized connection}
$\omega_\beta{}^\alpha$ which includes, besides the Lorentz (spin)
connection $\omega_L{}_\beta{}^\alpha= 1/4
\omega_L^{ab}\Gamma_{ab}{}_\beta{}^\alpha$, a tensorial part
$t_{1\beta}{}^\alpha= \omega_\beta{}^\alpha-
\omega_L{}_\beta{}^\alpha$
 constructed from the fields of
supergravity multiplet or their field strengths.
 Among the cases
where the generalized connection is reduced to the Lorentz
connection (${\cal D}=D$, $t_{1\alpha}{}^\beta =0$) is the simple,
$D=4$ $N=1$, supergravity. In $D=11$ supergravity the Lorentz
covariant part  $t_{1\alpha}{}^\beta$ of the generalized connection
$\omega_\beta{}^\alpha$ is constructed in terms of  the field
strength $F_{abcd}$ of the three--form gauge field $A_3$,
\begin{eqnarray}\label{CJSom}
 \omega_\beta{}^\alpha &=& {1\over 4}
\omega_L^{ab}\Gamma_{ab}{}_\beta^{\;\;\alpha}+ {i\over 18} e^a\,
\left( F_{ab_1b_2b_3} \Gamma^{b_1b_2b_3}+ {1\over 8}
F^{b_1b_2b_3b_4}
\Gamma_{ab_1b_2b_3b_4}{}\right)\,{}_\beta^{\;\;\alpha} \; . \qquad
\end{eqnarray}
The {\it necessary condition} for the existence of  Killing
spinors  is given by an  algebraic equation coming from the
integrability condition for (\ref{Killing}) ${\cal D}{\cal D}
\in_J{}^\alpha =0$. It has the suggestive form \cite{Duff03},
\begin{eqnarray}\label{eIR=0}
\in_{\check{J}}{}^\beta  {\cal R}_\beta{}^{\alpha}  = 0 \qquad
\Leftrightarrow \qquad {\cal D}{\cal D} \in_J{}^\alpha =0 \; ,
\qquad
\end{eqnarray}
in terms of the {\it generalized curvature}
\begin{eqnarray}\label{calR}
{\cal R}_\beta{}^{\alpha} = d\omega_\beta{}^{\alpha} - \omega
_\beta{}^{\gamma}\wedge  \omega_\gamma{}^{\alpha} \;
\end{eqnarray}
or curvature of generalized connection taking  values in the Lie
algebra of the so--called {\it generalized holonomy group}
\cite{Duff03,Hull03}. For $D=11$ supergravity the generalized
holonomy group has to be a subgroup of $SL(32,\mathbf{C})$
\cite{Hull03} (see \cite{Oscar+} for further discussion with
concrete solutions). The same is true for type IIB supergravity
\cite{P+T031}.

The r\^ole of BPS preons in the analysis of supersymmetric
supergravity solutions was discussed in \cite{BPS03}. The fact that
a BPS state $\vert BPS\; k \rangle$ preserving $k$ of $32$ (in
general $k$ of  $n$) supersymmetry can be considered as a composite
of $(32-k)\;$ (in general ($n-k$)) BPS preons, Eq. (\ref{k=npreon}),
is reflected, in the language of  supergravity  solutions,  by the
possibility of finding $\tilde{n}:=(32-k)$ spinors (spinor fields)
$\lambda_\alpha^r(x)$, $r=1, \, \ldots \, , \, \tilde{n}$ which are
orthogonal to the Killing spinors, $\in_{\check I}{}^\alpha(x)$,
${\check I}= 1, \ldots , k$, Eq. (\ref{Killing}), characterizing the
solution,
\begin{eqnarray}\label{eIlr=0}
 \in_{\check{J}} {}^\alpha(x)\, \lambda_\alpha{}^r(x) &=& 0 \, ,
 \qquad   \check{J}=1,\ldots,  k \, , \quad  r =1,\ldots, (32-k) \; .
\end{eqnarray}
Thus, BPS preonic spinors and Killing spinors provide an alternative
(dual) characterization of a $\nu=k/32$--supersymmetric solution
($\nu=k/n$ in general); either one can be used and, for solutions
with supernumerary supersymmetries \cite{pp>1/2}, the description
provided by BPS preons is clearly a more economic one. Moreover, the
use of both BPS preonic ($\lambda_\alpha{}^r$) and Killing
($\epsilon_J{}^\alpha$) spinors allowed us to develop \cite{BPS03} a
{\it moving G--frame} method, which may be useful in the search for
new supersymmetric solutions of  supergravity.

As a simplest application of the moving $G$--frame method let us present the general
expression for the generalized curvature (\ref{calR}) of the $k/32$--supersymmetric
solution of $D=11$ Cremmer--Julia--Scherk supergravity \cite{BPS03}. It is given by
\begin{equation}\label{kcalR=gl}
 {\cal R}_\alpha{}^\beta = G_r{}^s \, \lambda_{\alpha}{}^r\, w_s{}^\beta
+ \nabla B_r^{\check{I}}\, \lambda_{\alpha}{}^r \,
\in_{\check{I}}{}^\beta \; ,
\end{equation}
where $w_s{}^\beta$ is a set of $(32-k)$ spinors obeying
$w_s{}^\beta \lambda_\beta{}^r= \delta_s{}^r$ and forming,
together with the Killing spinor $\in_{\check{J}}{}^\alpha$,  the
nondegenerate matrix
\begin{eqnarray}\label{g}
g^{-1}{}_{(\beta)}{}^\alpha = \left( \matrix{ w_s{}^\alpha \cr
\in_{\check{J}}{}^\alpha } \right)\qquad \left( g_\alpha{}^{(\beta
)} =  \left( \matrix{\lambda_\alpha{}^s \; , \;
w_\alpha{}^{\check{J}}} \right)\right)
\end{eqnarray}
the moving G--frame matrix. Finally,
\begin{eqnarray} \label{Grs=}
 && G_r{}^s :=  (dA - A\wedge A)_r{}^s \; , \qquad
\nabla B_r^{\check{I}}:= dB_r^{\check{I}} - A_r{}^s \wedge
B_s^{\check{I}} \; ,
\end{eqnarray}
where $A_s{}^r$ and $B_r{}^{\check{I}}$ are $\tilde{n}\times
\tilde{n}\equiv (32-k)\times (32-k)$ and $\tilde{n}\times k \equiv
(32-k)\times k$ matrix valued one--forms which have to be fixed by
 the concrete $k/32$--supersymmetric solution. The
condition that the generalized holonomy group $H$ (as well as the
generalized structure group $G$) should be inside $SL(32)$, ${\cal
R}_{\alpha}{}^{\alpha}=0$, \cite{Hull03} implies that $G_r{}^r=0$
($A_r{}^r=0$).

On one hand, Eq. (\ref{kcalR=gl}) with $G_r{}^r=0$  provides an explicit expression for
the results of \cite{Hull03,P+T031} that the  generalized holonomies of
$k/32$--supersymmetric solutions of $D=11$ and of $D=10$ type IIB supergravity should
be
 $\; H\subset \; SL(32-k,\mathbf{R})$ $\subset\!\!\!\!\!\!\!\!\times
\mathbf{R}^{k(32-k)}$. In the light of the fact that, when fermions
vanish, $\psi_\mu^\alpha=0$,   all the free bosonic equations of the
11--dimensional supergravity as well as the Bianchi identities  for
the Riemann tensor and for the three--form gauge field strengths can
be collected in the following simple equation for generalized
curvature (\ref{calR}) \cite{GP02,BPS03}
\begin{equation}\label{EqCJS=0}
{\cal R}_{ab\,\alpha}{}^{\gamma}\; \Gamma^b{}_{\gamma}{}^{\beta}
=0\; ,
\end{equation}
we expect that the explicit form (\ref{kcalR=gl}) of the generalized
curvature ${\cal R}$ to  be useful in the search for new
supergravity solutions \footnote{The concise form of all the bosonic
equations for $D=11$ supergravity generalizing (\ref{EqCJS=0}) for
the case of nonvanishing fermions can be found in \cite{11DEq}.}.

In particular, the moving $G$--frame formalism might be useful to
settle  the question whether a BPS preonic solution preserving $31$
out of the $32$ supersymmetries exists in $D=11$ and/or $D=10$ type
IIB supergravities. Although this problem was addressed in
\cite{Duff03,Hull03,BPS03}, neither a solution with such a property
has been found  nor a statement forbidding an existence of such a
solution has been proved yet. However in \cite{BPS03} it was
observed that BPS preonic configurations do solve the equations of a
Chern--Simons like supergravity. This follows from the fact that the
generalized curvature of a BPS preon is nilpotent, ${\cal
R}_\alpha{}^\gamma \wedge {\cal R}_\gamma{}^\beta = 0$.  This
actually follows from the statement of \cite{Hull03} that the
generalized holonomy of (a hypothetical) $\nu= 31/32$ supersymmetric
solution  is a subgroup of $\mathbf{R}^{31}$. More explicitly,
according to \cite{BPS03} the generalized curvature for the preonic
($\nu= 31/32$) solution has the form
\begin{eqnarray}\label{calR=sl}
 && {\cal R}_\alpha{}^\beta = dB^I  \, \lambda_{\alpha}\; \in_I{}^\beta \;  \qquad
\end{eqnarray}
which, in the light of (\ref{eIlr=0}), implies the nilpotency not only for the two form
${\cal R}_\alpha{}^\beta$ but also for the tensor ${\cal R}_{ab\, \alpha}{}^\beta$. See
\cite{BPS01}  for a further discussion on a hypothetical preonic solution in $D=11$
Cremmer--Julia--Scherk supergravity.

\subsection{III.5. Superparticle model for BPS preon}

Interestingly enough, the point--like model for BPS preon
\cite{BPS01} is provided by the action that had been proposed in
\cite{BL98} before the notion of BPS preons was introduced. This
describes a superparticle in {\it tensorial superspace} (which was
called ''generalized superspace'' in \cite{BL98},  ''extended
superspace''  in \cite{JdA00} and enlarged superspace in
\cite{30/32}) with the bosonic body (\ref{Sn0}),
\begin{eqnarray}
\label{Sn} \Sigma^{({n(n+1)\over 2}|n)}\; & : & {\cal Z}^{{\cal
M}}= (X^{\alpha\beta}, \theta^\alpha)\; , \quad X^{\alpha\beta}=
X^{\beta\alpha}\; , \qquad   \alpha, \beta = 1,2, \ldots , n \; .
\end{eqnarray}
The action of ref. \cite{BL98} is the straightforward
supersymmetric generalization  of the bosonic functional
(\ref{p=0preon0}), which can be obtained by substituting the
pull--back $\hat{\Pi}^{\alpha\beta} \equiv d\tau
\hat{\Pi}_\tau^{\alpha\beta} = d\hat{X}^{\alpha\beta}(\tau) - i
d\hat{\theta}^{(\alpha}\,\hat{\theta}^{\beta )}(\tau)$ of the
supersymmetric Volkov--Akulov one--form
\begin{eqnarray}\label{PiS}
{\Pi}^{\alpha\beta} := d{X}^{\alpha\beta} - i
d{\theta}^{(\alpha}\,{\theta}^{\beta )}\;
\end{eqnarray}
of the $\Sigma^{({n(n+1)\over 2}|n)}$ superspace for
$d\hat{X}^{\alpha\beta}$ in (\ref{p=0preon0}),
\begin{eqnarray}
\label{p=0preon} S = {1\over 2} \int_{W^1} \lambda_\alpha
\lambda_\beta \hat{\Pi}^{\alpha\beta} \equiv {1\over 2} \int d\tau
\;  \, \hat{\Pi}_\tau^{\alpha\beta}(\tau)\,
\lambda_{\alpha}(\tau)\lambda_{\beta}(\tau) \; ,
\end{eqnarray}
The action (\ref{p=0preon}) is invariant under the global
supersymmetry transformations,
\begin{eqnarray}\label{susy}
\delta_{\epsilon} \hat{X}^{\alpha\beta}(\tau) = -
i\epsilon^{(\alpha}\hat{\theta}^{\beta )}(\tau)\, , \quad
\delta_{\epsilon} \hat{\theta}^{\alpha}(\tau)= \epsilon^{\alpha} \,
, \qquad \delta_{\epsilon} \lambda_{\alpha}(\tau)=0\; ,
\end{eqnarray}
and under $(n-1)$ local fermionic $\kappa$--symmetries
\begin{eqnarray}
\label{kappa=defb} i_{\kappa} \hat{\Pi}^{\alpha\beta}&=& 0 \quad
\Leftrightarrow \quad \delta_\kappa \hat{X}^{\alpha\beta} = 2i
\delta_\kappa \hat{\theta}^{(\alpha} \hat{\theta}^{\beta )} \; ,
\hspace{1cm}  \delta_\kappa \lambda_\alpha  =  0\; , \qquad  \\
\label{kappa=deff} \delta_\kappa \hat{\theta}^\alpha \,
\lambda_\alpha (\tau) &= & 0 \; \quad \Leftrightarrow \quad
\delta_\kappa \hat{\theta}^\alpha = \hat{u}_I{}^\beta (\tau)
\kappa^I(\tau)\; , \quad I=1, \ldots , (n-1) \; .
\end{eqnarray}
In (\ref{kappa=defb}) $\kappa^I(\tau)$ are the $(n-1)$ ($31$ for
$D=11$) Grassmann parameters of the $\kappa$--symmetry and
$\hat{u}_I{}^\beta (\tau)$ are $(n-1)$ ($31$) auxiliary spinor
fields  which are orthogonal to $\lambda_\alpha(\tau)$, Eq.
(\ref{uI=def0}) or Eq. (\ref{eIl=0}) with (\ref{e=u}). These can
be omitted from the consideration as one can use  the first
equation in (\ref{kappa=deff}), $\delta_\kappa \hat{\theta}^\alpha
\, \lambda_\alpha (\tau)= 0$, as the definition of the
$\kappa$--symmetry.

 Just the presence of $31$--parametric ($n$--parametric) $\kappa$--symmetry,
 allows one to treat the action (\ref{p=0preon}) as a model
for BPS preons: {\it the $\kappa$ symmetry of the worldvolume action
reflects the supersymmetry preserved by the ground state of the
point--like or extended object \cite{BKOP97,ST97}}. Indeed the
requirement of the Lorentz invariance (or more powerful $GL(n)$
invariance) of the ground  state leads to the conclusion that in
this state all fermions vanish, $\hat{\theta}^\alpha(\tau)=0$. Then,
as the fermionic coordinate function transforms both under the
supersymmetry and under the $\kappa$--symmetry,  $\delta
\hat{\theta}^{{\alpha}}= \varepsilon^\alpha +
\delta_{\kappa}\hat{\theta}^{{\alpha}}$, the invariance of the
ground state of the superparticle (\ref{p=0preon}) is defined by the
equation
\begin{eqnarray}\label{e=kkap}
\hat{\theta}^{{\alpha}} =0 \; \Rightarrow \qquad 0= \delta
\hat{\theta}^{{\alpha}}= \varepsilon^\alpha +
\delta_{\kappa}\tilde{\theta}^{{\alpha}}= \varepsilon^\alpha +
\hat{u}^\alpha_I \kappa^I(\tau)  \; .
\end{eqnarray}
Thus  the parameters of the symmetries preserving ground state
solution should obey
\begin{eqnarray}\label{SUSY=uIkI}
\hbox{{\small susy preserved  by  ground  state with
$\hat{\theta}^{{\alpha}}= 0$}:} \qquad & \varepsilon^\alpha = -
\hat{u}^\alpha_I \kappa^I\; . & \qquad
\end{eqnarray}
The extended object models for
BPS preons are provided by tensionless superbranes in
$\Sigma^{(528|32)}$ ($\Sigma^{(n(n+1)/2|n)}$) superspace
\cite{ZU,B02}.

\section{IV. Superfields and Supergravity in tensorial superspace}

\subsection{IV.1. Superfield generalization of the massless higher spin equations}

The superparticle models \cite{BL98} with the the properties of BPS
preon \cite{BPS01} were studied in the flat tensorial superspace
$\Sigma^{(n(n+1)/2|n)}$ \cite{BLS99,V01s,Dima03} and on the
$OSp(1|2n)$ supergroup manifold \cite{BLPS,Misha+,Dima03}. The
latter are the ''AdS--like'' version of tensorial superspace (see
 \cite{BLPS,V01s,Misha+,Dima03},  Sec. II.4 for a brief discussion
 and
\cite{BBdAST04} for $D=6,10$ generalization of this statement). The quantum state
spectrum of the preonic superparticle in $D=4$ contains a tower of conformal massless
fields of all possible 'helicities'; which can be described all together (see
\cite{BLS99,BBdAST04} and Sec. II.4) by the scalar bosonic and spinor (s-vector)
fermionic fields obeying Eqs. (\ref{ddb=0}) and (\ref{df=0}).

This spectrum is manifestly supersymmetric. Then the question
arises: is there any superfield generalization of these equations,
{\it i.e.} is there a superfield equation which collects the scalar
and spinor field in tensorial space and implies  Eqs. (\ref{ddb=0})
and (\ref{df=0}) on these fields? The answer on this question is
affirmative. As it was shown in \cite{BPST04},  such a superfield
equation does exist and has the form
\begin{equation}\label{hsSEq}
D_{[\alpha}D_{\beta ]} \Phi (X, \theta ) = 0 \, ,
\end{equation}
where
 $D_\alpha=
\partial /\partial \theta^\alpha + i \theta^{\beta}
\partial_{\beta\alpha}$ is the flat Grassmann  covariant derivative in
the flat tensorial superspace $\Sigma^{(n(n+1)/2|n)}$, (\ref{Sn}),
$(\{D_\alpha , D_\beta \}=2i\partial_{\alpha\beta})$. Eq.
(\ref{hsSEq}) sets to zero all the higher components
$\phi_{\alpha_1\cdots\alpha_i}(X)$, $i\geq 2$, of the scalar
superfield $\Phi(X,\theta)=b(X)+f_\alpha(X)\,\theta^\alpha+
\sum_{i=2}^{n}\phi_{\alpha_1\cdots\alpha_i}(X)\,\theta^{\alpha_1}\cdots\theta^{\alpha_i}$,
thus reducing it to the form
\begin{equation}\label{scalar}
\Phi (X^{\alpha\beta} ,
\theta^\gamma)=b(X)+f_\alpha(X)\,\theta^\alpha\; ;
\end{equation}
it also imposes on the surviving components the dynamical equations
(\ref{ddb=0}) and (\ref{df=0}). \footnote{One can also collect the
same field content inside a spinor superfield $\Psi_\alpha$, but
this should be subject to a set of two equations,
$D_{[\alpha}\Psi_{\beta ]} (X, \theta )= 0 $ and
$\partial_{\alpha[\beta} \Psi_{\gamma ]} (X, \theta ) =0$
\cite{BPST04}.}  The generalization of the ``preonic equation''
(\ref{(P-ll)Phi(X,l)}) has the form \cite{BPST04}
\begin{equation}\label{preonsusy}
(D_{\alpha}D_{\beta }+\lambda_\alpha\lambda_\beta)\, \Phi (X\, ,
\, \theta \, , \, \lambda ) = 0 \,.
\end{equation}
Its antisymmetric part gives Eq. (\ref{hsSEq}) while the symmetric
part produces  Eq. (\ref{(P-ll)Phi(X,l)}).

The $AdS$ generalization of Eq. (\ref{hsSEq}) reads
\begin{eqnarray}\label{SeqOSp}
\left(\nabla_{[\alpha} \nabla_{\beta]} + i{\varsigma\over 4}
C_{\alpha\beta} \right) \Phi(X,\theta)=0 \; ,
\end{eqnarray}
where $\nabla_{\alpha}$ are spinorial covariant derivatives on the
$OSp(1|n)$ supergroup manifold obeying the superalgebra
(\ref{OSpff}), ((\ref{OSpbf}), (\ref{OSpbb}).  The equation
generalizing (\ref{preonsusy}) for the case of $OSp(1|n)$ supergroup
manifold reads
 can be found in \cite{BPST04} where it was also discussed the way of derivation of
 (\ref{preonsusy}) from the equation for Clifford  superfield
 wave function which appeared in the quantization \cite{BLS99} of the
preonic superparticle \cite{BL98}, Eq. (\ref{p=0preon})  with the
conversion method.

\subsection{IV.2. Superfield supergravity in tensorial superspace}

When one considers the standard superparticles and superbranes, the
natural starting point is also an action in flat superspace. Then
one finds \cite{BST87,strKsg} that considering the natural
generalization of the model in curved superspace and assuming the
existence of a smooth flat superspace limit one arrives at the
requirement that curved  superspace has  to satisfy the supergravity
constraints. For higher dimensional $D>6$ superspaces (and also for
the extended, $N> 2$, superspaces in $D=4$) these are the {\it
on--shell} supergravity constraints which contain all the dynamical
equations of motion among their consequences. This is not the case
in $D=4$ $N=1$ superspace where one arrives at off-shell constraints
which do not imply dynamical equations of motion (see refs. in
\cite{BI03} which is devoted to a complete Lagrangian description of
the supergravity---superstring interaction); in a simpler $D=3$ and
$D=2$ cases the supergravity is not dynamical.

It is natural to ask what are the generalized supergravity
constraints which might appear from the consistency requirement for
a preonic model in a curved tensorial superspace. Such a
supergravity in a curved $\Sigma^{(528|32)}$ superspace may be
interesting in an M--theoretical perspective, while the models in
$\Sigma^{(n(n+1)/2|n)}$ with $n=4,6$ and $10$ [$\Sigma^{(10|4)}$,
$\Sigma^{(36|8)}$ and $\Sigma^{(136|16)}$] could provide a basis for
interacting higher spin theories. One might even hope that such a
tensorial supergravity could itself provide an interacting higher
spin theory; however, as shown in \cite{BPST04}, this is not the
case, at least when the supergravity with $SL(n)$ or $GL(n)$
holonomy groups are considered.

The natural generalization of the point like preonic action
(\ref{p=0preon}) for the case of curved tensorial superspace
$\Sigma^{(n(n+1)/2|n)}$ reads \cite{BPST04}
\begin{eqnarray}
\label{p=0prSG} S = {1\over 2} \int_{W^1} \lambda_\alpha
\lambda_\beta \hat{E}^{\alpha\beta} \equiv {1\over 2} \int d\tau
\;  \, \hat{E}_\tau^{\alpha\beta}(\tau)\,
\lambda_{\alpha}(\tau)\lambda_{\beta}(\tau) \; ,
\end{eqnarray}
where $\hat{E}^{\alpha\beta}:= d\tau \,
\hat{E}_\tau^{\alpha\beta}= d\tau \, \partial_\tau \hat{{\cal
Z}}^{\cal M}\; E_{\cal M}{}^{\alpha\beta}(\hat{{\cal Z}})$ is the
pull--back to the worldline $W^1$ of the bosonic supervielbein
form ${E}^{\alpha\beta}:= d{{\cal Z}}^{\cal M}\; E_{\cal
M}{}^{\alpha\beta}({{\cal Z}})$ of the curved  tensorial
superspace ${\Sigma}^{(n(n+1)/2|n)}$  (\ref{Sn})
 with
supervielbein
\begin{eqnarray}\label{E528}
{E}^{\cal A} = (\, {E}^{{\alpha}{\beta}}\, , \,
 E^{{\alpha}})\, = d{{\cal Z}}^{\cal M} E_{\cal M}{}^{\cal A}({\cal Z}) \; , \qquad {\alpha}\, ,
 \, {\beta} = 1 \, , \, \ldots \, , \, n \;
\end{eqnarray}
including also  $n$ fermionic one forms $E^{\alpha}$ whose
pull--backs $\hat{E}^{\alpha}$ do not enter Eq. (\ref{p=0prSG}).

The supergravity in tensorial superspace should be the theory of the
supervielbein superfields in (\ref{E528}). However, to make the
formalism covariant one also introduces in superfield supergravity
the connection taking values in a structure group of the superspace.
For the usual superspace the structure group is the Lorentz group
which in the flat superspace appears as a global symmetry of the
superparticle action. With this in mind, and taking into attention
that the flat superspace preonic superparticle action
(\ref{p=0preon}) is invariant under $GL(n)$ group, one finds natural
to consider $GL(n)$ as the structure group of
 tensorial superspace. Hence,
 by analogy with the conventional spin connection of general
relativity and the standard supergravity, the $GL(n)$ connection was
introduced in \cite{BPST04},
\begin{equation}
\label{Om} \Omega_{\beta}{}^{\alpha}:= d{\cal Z}^{{\cal M}}
\Omega_{{{\cal M}} \beta}{}^{\alpha} \equiv E^{{\cal A}}
\Omega_{{\cal A}\beta}{}^{\alpha}\, .
\end{equation}
The torsion 2-forms and the curvature of the $GL(n)$ connection
were defined by
\begin{eqnarray}\label{Tb=def}
T^{\alpha\beta}& := & {\cal D} E^{\alpha\beta} \equiv
dE^{\alpha\beta} - E^{\alpha\gamma}\wedge \Omega_{\gamma}{}^{\beta
}-E^{\beta\gamma}\wedge
\Omega_{\gamma}{}^{\alpha }\,, \\
\label{Tf=def} T^{\alpha}\; & := & {\cal D} E^{\alpha} \equiv
dE^{\alpha} - E^{\beta}\wedge \Omega_{\beta}{}^{\alpha} \,
\\
 \label{R=def}
{\cal R}_{\beta}{}^{\alpha}\! & := & d\,
 \Omega_{\beta}{}^{\alpha} -
\Omega_{\beta}{}^{\gamma}\wedge \Omega_{\gamma}{}^{\alpha} \,.
\end{eqnarray}

The requirement of preservation of the $\kappa$--symmetry of the
superparticle (\ref{p=0prSG}) imposes the constraints $T_{\gamma \;
\delta }{}^{\alpha\beta} \propto \delta_{\gamma}{}{}^{(\alpha} \;
\delta_\delta{}^{\beta )}$, $T_{\gamma\gamma^\prime \; \delta
}{}^{\alpha\beta} \propto \delta_{(\gamma}{}{}^{\alpha} \;
t_{\gamma^\prime) \delta}{}^{\beta}$,
 on the bosonic torsion (\ref{Tb=def}),
$T^{\alpha\beta}:= {1\over 2} E^{{\cal D}} \wedge  E^{{\cal C}} \, T_{{\cal C}{\cal
D}}{}^{\alpha\beta}$. Then imposing the conventional constraints, which fix the freedom
in redefinition of the basic superfields, and studying the Bianchi identities one finds
the following complete expressions for the torsion and curvature two--forms
\cite{BPST04}
\begin{eqnarray}\label{Tb=AdS}
T^{\alpha\beta}& = & -i E^{\alpha}\wedge E^{\beta} + 2
E^{\gamma(\alpha}\wedge E^{\beta)\delta } R_{\gamma\delta } \;,
\\ \label{Tf=AdS} T^{\alpha}& = & 2 E^{\alpha\beta}\wedge
E^{\gamma} R_{\beta\gamma } + E^{\alpha\beta}\wedge
E^{\gamma\delta } U_{\beta{\gamma\delta }}
\,, \\
\label{R=AdS} {\cal R}_\beta{}^\alpha &= & i E^{\gamma \delta}
\wedge E^\alpha U_{\beta{\gamma\delta }} - E^{\alpha \gamma}
\wedge E^\delta (F_{\delta {\beta\gamma}}+ {\cal D}_\delta
R_{\beta\gamma}) - \nonumber \\ &&  \hspace{3cm} -\,  E^{\alpha
\gamma} \wedge E^{\delta\epsilon} ({\cal D}_{(\beta} U_{\gamma )
{\delta\epsilon}} + {\cal D}_{\delta\epsilon} R_{\beta\gamma}) \,.
\quad
\end{eqnarray}
Here  $R_{\gamma\delta }({\cal Z})= - R_{\delta\gamma }({\cal Z})$
and $U_{\alpha \, {\beta
    \gamma}}({\cal Z})= U_{\alpha \, {\gamma\beta}}({\cal Z})$ are `main'
superfields which are related by the equations
\begin{eqnarray}\label{DU=DR}
{\cal D}_{[\alpha }U_{\beta] {\gamma\delta}} &=& - {\cal
D}_{\gamma\delta} R_{\alpha\beta} \;,
\\ \label{DU=DF}
{\cal D}_{(\alpha}U_{\beta)\, {\gamma\delta}} &=& - i {\cal
D}_{(\gamma} F_{\delta )\; \alpha\beta} \; , \qquad F_{\alpha
{\beta\gamma}}= 2i U_{(\beta {\gamma )\alpha}} - iU_{\alpha
  {\beta\gamma}} - 2 {\cal D}_{(\beta} R_{\gamma ) \alpha}\, , \quad \\
\label{5/2} {\cal D}_{\alpha \beta} U_{\gamma\delta \, \sigma} &-&
{\cal D}_{\delta \sigma} U_{\gamma \alpha \beta} +2  U_{\gamma
\alpha (\sigma } R_{\delta)\beta } + 2U_{\gamma\beta(\sigma }
R_{\delta)\alpha } = 0 \, .
\end{eqnarray}

Setting the main superfields to zero, $R_{\gamma\delta }({\cal
Z})= 0$, $U_{\alpha \, {\beta
    \gamma}}({\cal Z})= 0$
    and ignoring the trivial $GL(n)$ connection
    (setting $\Omega_{\alpha}{}^{\beta}=0$) one
reduces the constraints to the Maurer--Cartan equations of flat
tensorial superspace $\Sigma^{(n(n+1)/2|n)}$ with the solution
\begin{eqnarray}\label{flatV}
R_{\gamma\delta }(Z)= 0\, , \quad U_{\alpha \, {\beta
    \gamma}}(Z)= 0\, \quad \Rightarrow \quad
E^{\alpha\beta}= \Pi^{\alpha\beta} \; , \quad E^{\alpha}=
d\theta^{\alpha} \;  . \end{eqnarray}
 On the other hand, setting $R_{\gamma\delta }=
\varsigma C_{\gamma\delta}$ and $U_{\alpha \, {\beta
    \gamma}}(Z)= 0$ one arrives in the Maurer--Cartan equations
    of the $OSp(1|2n)$ supergroup
\begin{eqnarray}\label{OSPMC}
d {\cal E}^{\alpha\beta} & = &- i {\cal E}^{\alpha} \wedge {\cal
E}^{\beta} - \zeta {\cal E}^{\alpha\gamma} \wedge
{\cal E}^{\delta\beta} C_{\gamma\delta}  \; , \nonumber \\
d {\cal E}^{\alpha}& = &  - \zeta  {\cal E}^{\alpha\gamma} \wedge
{\cal E}^{\delta} C_{\gamma\delta}  \; .
\end{eqnarray}
In both cases the curvature is equal to zero which allows
    one to
    gauge away the trivial $GL(n)$ connections
    $\Omega_{\alpha}{}^{\beta}=0$.

In the superspace subject to the constraints (\ref{Tb=AdS}) the
preonic superparticle action
 possesses the gauge invariance under the local fermionic
 $\kappa$--symmetry ({\it cf.} (\ref{kappa=defb}),
 (\ref{kappa=deff}))
\begin{eqnarray}\label{kappaGb1}
i_\kappa E^{{\alpha \alpha^\prime}}:= \delta_\kappa Z^{{M}} E_{
{M}}^{{\alpha \alpha^\prime}} =0 \; , \quad
 i_\kappa E^{\alpha} :=  \delta_\kappa Z^{{M}}
E_{ {M}}^{\alpha} = u^{\alpha}_I\kappa^I(\tau)\; , \quad
\end{eqnarray}
where $u^{\alpha}_I$ is defined by Eq. (\ref{uI=def0}),  and under
the $b$--symmetry transformations ({\it cf.} (\ref{vbX00}))
\begin{eqnarray}\label{b-symS}
i_b E^{\alpha\alpha^\prime} := \delta_b Z^{{M}} E_{ {M}}^{{\alpha
\alpha^\prime}} = u^{\alpha}_I u^{\alpha^\prime}_J b^{IJ} (\tau)\;
,   \quad
 i_b E^{\alpha} :=  \delta_b Z^{{M}}
E_{ {M}}^{\alpha} =0 \; .
\end{eqnarray}

One can ask whether the scalar superfield equation (\ref{hsSEq})
allows for a consistent generalization to the curved tensorial
superspace. It does and the desired generalization reads
\begin{equation}\label{DDR}
{\cal D}_{[\alpha}{\cal D}_{\beta]}\,\Phi={i\over
2}\,R_{\alpha\beta}\,\Phi\;
\end{equation}
and is consistent when the holonomy group of the tensorial superspace is restricted to
be $SL(n)$ (which means that the curvature tensor is traceless ${\cal
R}_\alpha{}^\alpha=0$) \cite{BPST04}.

The first impression might be that the tensorial supergravity
defined by the constraints (\ref{Tb=AdS})--(\ref{R=AdS})  should
contain a huge number of extra nonphysical fields. However this is
not the case. As it is shown in \cite{BPST04} the general solution
of the tensorial supergravity constraints contains only two classes
of superspaces:  the {\it superconformally flat superspaces} and the
superspaces superconformally related to the $OSp(1|n)$ supergroup
manifold. The superconformally flat superspaces are described by
\begin{eqnarray}\label{e}
& E^{\alpha\beta}= e^{{2W(Z)}\over n}\,\Pi^{\alpha^\prime
\beta^\prime}\, L_{\alpha^\prime}^{~\alpha}(Z)\; L_{
\beta^\prime}^{~\beta}(Z)\; , \qquad E^{\alpha} =e^{{W(Z)}\over
n}\,(d\theta^{\alpha^\prime}-i\Pi^{\alpha^\prime
\beta^\prime}\,D_{\beta^\prime}\,W) \,
L_{\alpha^\prime}^{~\alpha}(Z)\; , \nonumber
\\ \label{omega}
& \Omega^{~\alpha}_\beta= {1\over n} dW\,\delta_\beta{}^\alpha
-L^{-1}{}_{\beta}{}^{\beta '}\,\left[d\theta^{\alpha'} \,D_{\beta
'} W+\Pi^{\alpha '\gamma} (D_{\gamma\beta '}W+{i\over 2}D_\gamma
W\,D_{\beta'} W)\right]L_{\alpha'}^{~\alpha}\; ,  \;
\end{eqnarray}
where $L_{\beta}^{~\alpha}(Z)$ is a matrix of local $SL(n)$
transformations which together with $exp\{ {W(Z)}/ n\}$ form a
$GL(n)$ matrix $G_{\beta}^{~\alpha}=L_{\beta}^{~\alpha}exp\{
{W(Z)}/ n\}$. The extraction of the scaling factor allows to apply
Eqs. (\ref{e}) to supergravity with $SL(n)$ structure group.

Working with the structure group $GL(n)$ (which does not forbid reduction of the {\it
holonomy} group down to its subgroup $SL(n)$), one can obtain all superspaces
superconformally--related to the $OSp(1|n)$ supergroup manifold by making first the
following ''generalized super--Weyl transformations'' \cite{BPST04}
\begin{eqnarray}\label{Wa}
E^{\alpha\beta}&=& {\cal E}^{\alpha\beta}\,,\qquad E^{\alpha}\; =
{\cal E}^{\alpha}+ {\cal E}^{\alpha\beta}W_\beta \qquad \nonumber
\\
\Omega_{\beta}^{\; \; \alpha} &=&  -i{\cal E}^{\alpha}W_\beta
-{\cal E}^{\alpha\gamma}({\nabla}_{\gamma}\,W_\beta+iW_\gamma
W_\beta)\,, \quad
\end{eqnarray}
of the $OSp(1|n)$  supervielbein $({\cal E}^{\alpha\beta}\; , \;
{\cal E}^{\alpha})$, Eq. (\ref{OSPMC}),  with $W_\gamma=
-i{\nabla}_{\gamma}W$  and then performing a  $GL(n)$ ``rotation'',
if needed. In (\ref{Wa}) ${\nabla}_{\gamma}$ is the $OSp(1|n)$
covariant derivative, $d= {\cal E}^{\alpha\beta}
{\nabla}_{\alpha\beta}+ {\cal E}^{\alpha}{\nabla}_{\alpha}$.  The
flat tensorial superspaces $\Sigma^{(n(n+1)/2|n)}$ can be recovered
in $\varsigma =0$ limit.

The fact that superconformally flat and $OSp(1|n)$ related
superspaces provide the general (modulo topological subtleties)
solution of the tensorial supergravity  constraints
(\ref{Tb=AdS})--(\ref{R=AdS}) implies that the main superfields can
always be expressed by
\begin{eqnarray}\label{Rosp}
R_{\alpha\beta}=-{\varsigma\over 2}C_{\alpha\beta}+{\cal
D}_{[\alpha}\,W_{\beta]}+ {i\over 2}\,W_\alpha W_\beta\,, \qquad
U_{\alpha\beta\gamma}=-{\cal
D}_{\beta\gamma}\,W_{\alpha}+W_{(\gamma} \,{\cal
D}_{\beta)}\,W_{\alpha}\,.
\end{eqnarray}
One can check that the holonomy group of the superspace reduces to
$SL(n)$ ({\it i.e.} that ${\cal R}_\alpha{}^\alpha=0$) when
$W_{\alpha} = -i{\cal D}_{\alpha}W$. The 'super--Weyl
transformations' (\ref{OSPMC}) with $W_{\alpha} \not= -i{\cal
D}_{\alpha}W$ result in the connection with $GL(n)$ holonomy. In the
$OSp(1|n)$ covariant derivatives the main superfields of the
superspace with $SL(n)$ holonomy group read
\begin{eqnarray}\label{DDRWo}
& R_{\alpha\beta}=\, i\,e^{-{{2W}\over n}}\,\left[\,
i{\varsigma\over 2}C_{\alpha\beta}+{ \nabla}_{[\alpha}{
\nabla}_{\beta]}\,W+{1\over 2}\,{ \nabla}_\alpha W\, {
\nabla}_\beta W\right]\,, \quad
\\ \label{UWfo}
& U_{\beta
 {\gamma\delta}}=e^{-{{3W}\over n}}
\left[- i {\nabla}_{\gamma\delta} { \nabla}_{\beta}
 W+
\nabla_{(\gamma }W\,\nabla_{\delta)}\nabla_{\beta}W\right]\,.
\end{eqnarray}

One can make the (seemingly important) observation that, formally,
putting in (\ref{DDRWo}) $R_{\alpha\beta}= - {\varsigma\over
2}C_{\alpha\beta}\, e^{(1+4/n)\, W/2 }$ one finds an equation
\begin{eqnarray}\label{Req-}
& \nabla_{[\alpha} \nabla_{\beta]} W + {1\over 2}
\nabla_{\alpha}W\, \nabla_{\beta} W = - {i\varsigma\over
2}C_{\alpha\beta} \left(1- e^{-{W\over 2}}\right)\;
\end{eqnarray}
for the scalar superfield $W$.  However, first one observes that, after the field
redefinition $W= 2\, \mathrm{ln} \left({\Phi + a\over a}\right)$ (with $a>0$) this
reduces to the scalar superfield equation (\ref{SeqOSp}) on the $OSp(1|n)$ supergroup
manifold. Moreover, this does {\it not} imply a nontrivial embedding of even the free
scalar superfield (and of the higher spin theories) in tensorial supergravity. The
reason can be traced to the super--Weyl invariance of both the constraints, Eqs.
(\ref{Tb=AdS})--(\ref{R=AdS}), and the scalar superfield equation in supergravity
background, Eq. (\ref{DDR}). Thus one may use (\ref{Wa}) as a field redefinition
(leaving the constraint invariant) pass form the superconformally--$OSp(1|n)$--related
geometry to the rigid $OSp(1|n)$ supergroup manifold.

In other words, like the $D=3$ $N=1$ Poincar\'e and $AdS$ supergravities, the
supergravity in tensorial superspace is shown to be nondynamical: the general solution
of its constraints is given by superconformally flat and $OSp(1|n)$ related superspaces
which may be reduced to the rigid $\Sigma^{(n(n+1)/2|n)}$ or $OSp(1|n)$ superspaces by
the (super)field redefinition (Eq. (\ref{Wa}) plus $GL(n)$ transformations or,
equivalently, Eqs. (\ref{omega})) \cite{BPST04}.

This implies, in particular, that to proceed with the search for $D=4, 6, 10$
interacting higher spin theories on the basis of curved tensorial superspaces
$\Sigma^{(n(n+1)/2|n)}$ with $n=4,8,16$  one has {\it to extend} the tensorial
superspace rather than to restrict it, as it might be expected.  On the other hand,
such an extension looks natural in the light of the existing results on interacting
higher spin theories \cite{Misha0304b,Vasiliev89}, as these imply the necessity of
doubling of the auxiliary variables. Such auxiliary  variables responsible for the spin
degrees of freedom can be chosen to be spinors or antisymmetric tensors ($y^{[\mu\nu
]}$ for $D=4$). Hence a natural candidate for the variables to use for the extension of
the tensorial superspace $X^{(\alpha\beta )}, \theta^\alpha)$ ($=(x^{\mu}, y^{\mu\nu},
\, , \theta^\alpha)$ for $D=4$) in a search for consistent interacting higher spin
theories are the bosonic spinors $\lambda_\alpha$ which are used to define the notion
of BPS preon and are present in the action \cite{BL98} for the ``preonic
superparticle''.

The study of supergravity and super--Yang--Mills theories in tensorial superspace
enlarged by additional bosonic spinors is an interesting subject for future study.
Another interesting direction is an M--theoretic use of the superconformally flat
and $OSp(1|32)$--related superspaces in an M--theoretical context. This might be
related with the direction described in the contribution of Jos\'e A. de Azc\'arraga to
this volume \cite{JdA04}.

\section{Concluding remarks}

In this contribution we made a brief review of the notion of BPS preon, both in its
original $D=11$ image as hypothetical constituents of M-theory \cite{BPS01} and in its
natural generalization to arbitrary dimension $D$ \cite{30/32,BPS03}. Actually the
definition of BPS preon possesses a wider $GL(n)$ symmetry and, thus is rather
characterized by the number $n$ of possible values of the spinor (or 's--vector'
\cite{V01s}) index $\alpha$ than by the number $D$ related to an invariance under a
subgroup $SO(t,D-t)\subset GL(n)$. For $n=4,8,16$ cases the BPS preon may be identified
with the tower   of massless higher spin fields in $D=4, 6$ and $10$. This can be
established
 by quantization \cite{BLS99} of the ``preonic
superparticle model''
 \cite{BL98} which, interestingly enough,
had been carried out some times before the notion of BPS preon was introduced in
\cite{BPS01}. The present treatment  \cite{30/32,BPS03,B02} of the results of
\cite{BLS99,BL98} in terms of BPS preon notion \cite{BPS01} is justified by a search
for a universal language which might
 provide a bridge between (essentially) eleven--dimensional M--theory
 and the higher spin theories in $D=4,6,10$.

We have also reviewed the ``preonic superparticle'' action of \cite{BL98} bringing us
to the tensorial superspace, as well as the r\^ole of BPS preons in the classification
and study of supergravity solitons \cite{BPS03}, the concise superfield description of
the higher spin theories \cite{BPST04} and the results of the study of supergravity in
tensorial superspace \cite{BPST04}.

Actually, in the light of the algebraic classification of the M--theory BPS states
proposed in \cite{BPS01}, the possibility of treating BPS preons as constituents of
M--theory is a bit more than conjecture; a conjecture concerns rather a usefulness of
such a treatment. One might express doubts on such a usefulness arguing that
 the symmetry of the BPS preon is too high to
describe the M-theory physics. However, its identification with higher spin theory in
lower dimensions suggests an answer. Higher spin theories were (and are) conjectured
(see \cite{Vasiliev89} and {\it e.g.} \cite{Dima04}) to be related to the ``symmetric''
phase of string theory characterized by an enhanced symmetry whose spontaneous
breaking should reproduce the complete string theory. In the same way one may
conjecture that the $GL(32)$--invariant (actually $OSp(1|64)$--invariant) description
provided by the BPS preons corresponds to a symmetric phase of M--theory, while the
complete description of M--theory might require the spontaneous breaking of these
$OSp(1|64)$ symmetry.

\begin{theacknowledgments}
The author is grateful  to Jos\'e A. de Azc\'arraga, Jurek Lukierski, Dmitri Sorokin
for discussions and useful comments. He thanks M. Ro\v cek, W. Siegel, C. Vafa and
other organizers of the Simons Workshop in Mathematics and Physics "Superstrings and
Topological Strings" for the conversations and hospitality in Stony Brook. The
contribution is based on work done in collaboration with J. A. de Azc\'arraga, J.
Lukierski, D. Sorokin, J. Izqierdo, M. Pic\'on, O. Varela and, more recently, P. Pasti
and M. Tonin, which is acknowledged with great pleasure. This work has been partially
supported by the research grants BFM2002-03681 from the Ministerio de Educaci\'on y
Ciencia and from EU FEDER funds, N383 from the Ukrainian State Fund for Fundamental
Research and by the {\rm Grupos 03-124} grant from the Generalidad Valenciana.

\end{theacknowledgments}

\end{document}